\documentclass[reprint,amsmath,amssymb,aps,pre,superscriptaddress]{revtex4-2}
\usepackage{amssymb,amsfonts}
\usepackage{mathrsfs}
\usepackage{amsmath}
\usepackage[svgnames]{xcolor}
\usepackage{framed}
\usepackage{tcolorbox}
\usepackage{lipsum}
\usepackage{lmodern}
\usepackage{tcolorbox}
\usepackage{tikz}
\usepackage{graphicx}
\usepackage[colorlinks=true]{hyperref}
            
\begin{document}

\title{Odd Diffusion in Three-Dimensional Isotropic Media}

\author{Viola Zixin Zhao}\email{zhaozixin@berkeley.edu}
\affiliation{Leinweber Institute for Theoretical Physics, University of California, Berkeley, CA, 94720-7300, USA}
\affiliation{Department of Physics University of California, Berkeley, CA, 94720-7300, USA}
\author{Andr\'{e}s Franco Valiente}\email{andresfranco@berkeley.edu}
\affiliation{Leinweber Institute for Theoretical Physics, University of California, Berkeley, CA, 94720-7300, USA}
\affiliation{Department of Physics University of California, Berkeley, CA, 94720-7300, USA}
\affiliation{Theoretical Physics Group, Lawrence Berkeley National Laboratory Berkeley, CA 94720-8162, USA}
\author{David T. Limmer}\email{dlimmer@berkeley.edu}
\affiliation{Department of Chemistry, University of California, Berkeley, CA, 94720-7300, USA}
\affiliation{Kavli ENSI, University of California, Berkeley, CA, 94720-7300, USA}
\affiliation{Chemical Sciences Division, Lawrence Berkeley National Laboratory Berkeley, CA 94720-8162, USA}
\affiliation{Material Sciences Division, Lawrence Berkeley National Laboratory Berkeley, CA 94720-8162, USA}

\begin{abstract}
Odd diffusion is a hallmark of chiral active matter, generating currents
transverse to density gradients. Existing theories rely on a linear antisymmetric transport coefficient that exists only in two dimensions, raising the question of whether odd diffusion can occur in isotropic three-dimensional systems. Here we show that such transport is possible
through a nonlinear constitutive law.
Symmetry considerations reveal that the three-dimensional Levi-Civita tensor
permits a leading order isotropic odd current at second order in the density gradient expansion and only in multicomponent systems. The resulting transport generates boundary-driven
rotational currents, finite vorticity, and enstrophy despite the absence of
external torques or preferred directions. We show how such a constitutive law derives from
a microscopic model of particles interacting through nonreciprocal three-body
forces using the Dean--Kawasaki coarse-graining procedure. These results
establish a minimal framework for odd transport in isotropic three dimensions.
\end{abstract}

\maketitle

\section{Introduction}

Transport coefficients that are antisymmetric under exchange of spatial indices have emerged as a defining feature of nonequilibrium chiral matter. Examples include odd viscosity and mobility in fluids, odd elasticity in solids, and odd diffusion \cite{scheibner2021odd,hargus2021odd,poggioli2023odd,banerjee2017odd}. Unlike their equilibrium counterparts, these transport coefficients do not derive from a free-energy functional and can generate currents perpendicular to thermodynamic forces in violation of Onsager reciprocity \cite{markovich2024nonreciprocity}. As a consequence, they provide a mechanism for sustained rotational motion and dissipationless energy transfer in systems maintained away from equilibrium \cite{poggioli2023emergent,alsallom2026origin,yang2020robust,siebers2024collective,soni2019odd,fruchart2023odd,metzger2026equation}. Their theoretical description challenges conventional ways to formulating hydrodynamic theories that rely on local equilibrium assumptions. Here we introduce a novel mechanism for odd transport, and describe its origin and hydrodynamic implications. 

Among documented odd transport phenomena, odd diffusion has attracted particular attention as a minimal realization of parity-breaking transport. In two spatial dimensions, the flux $\mathbf{J}$ of a conserved scalar density $\rho$ can contain an antisymmetric contribution\cite{hargus2021odd}
\begin{equation}
J_i=-D\partial_i \rho-D_{\rm odd}\epsilon_{ij}\partial_j\rho,
\end{equation}
where $\epsilon_{ij}$ is the two-dimensional Levi-Civita tensor indicating antisymmetry among spatial indices $i$ and $J$ where Einstein summation notation will be used throughout. The constants $D$ and $D_{\rm odd}$ are the regular and odd diffusion coefficients. The latter term rotates density gradients by ninety degrees, producing circulating currents without altering the underlying diffusive relaxation. Odd diffusion has been predicted and observed in a variety of active and chiral systems and is now recognized as part of a broader class of transport phenomena unique to systems that break time reversal symmetry \cite{kalz2022collisions,hargus2025odd,langford2025phase,hargus2025flux,vega2022diffusive}.

A fundamental question is whether isotropic odd diffusion can exist in three dimensions. At first sight, the answer appears to be negative. The constitutive law above relies on the existence of an isotropic antisymmetric rank-two tensor, an object that exists only in two dimensions. In three dimensions rotational invariance forbids such a tensor, suggesting that odd diffusion is intrinsically a lower-dimensional phenomenon. This conclusion parallels similar symmetry arguments that restrict the form of Hall-like responses in isotropic media \cite{von202040}.

In this work, we show that this obstruction is not fundamental. While isotropic linear odd diffusion is indeed forbidden in three dimensions, nonlinear couplings between multiple conserved fields allow a distinct form of odd transport that remains fully isotropic. Specifically, we demonstrate that the three-dimensional Levi-Civita tensor permits a symmetry-allowed current of the form
\begin{equation}
J_i^A=-D\partial_i \rho^A
-\omega \epsilon_{ijk} \epsilon^{ABC} (\partial_j\rho^B) (\partial_k\rho^C),
\label{eq:maincurrent}
\end{equation}
where $A,B,C$ label distinct conserved species. This constitutive law is the lowest-order isotropic odd transport term permitted by symmetry in three dimensions, and introduces a novel transport coefficient $\omega$ as a nonlinear odd diffusivity.

The resulting transport generated from Eq. ~\ref{eq:maincurrent} differs qualitatively from conventional odd diffusion. Because the nonlinear odd current is divergence-free in the bulk, it does not modify the steady-state density profiles of the conserved fields. Instead, it generates rotational currents whose structure is controlled entirely by boundary conditions. These currents possess finite vorticity and enstrophy despite the absence of globally imposed torques, preferred directions, or confining geometries. Odd transport in three dimensions therefore emerges not through bulk modification of diffusive relaxation, but through a fundamentally boundary-driven mechanism.

Beyond establishing the symmetry structure of three-dimensional odd diffusion, we derive a microscopic realization based on nonreciprocal three-body interactions between multiple particle species. Starting from stochastic particle dynamics and employing a Dean--Kawasaki procedure \cite{illien2025dean,irving1950statistical,kawasaki1994stochastic,dean1996langevin}, we show that the nonlinear constitutive law of Eq.~\ref{eq:maincurrent} emerges naturally at hydrodynamic scales. This construction provides a concrete route toward realizing isotropic three-dimensional odd transport in active and driven multicomponent systems. Whereas previous theories identified odd diffusion as a phenomenon tied to two-dimensional geometry, we show that odd transport persists in isotropic three dimensions once nonlinear multicomponent couplings are included. The work therefore extends the scope of odd hydrodynamics beyond its conventional setting and suggests new opportunities for realizing rotational nonequilibrium transport in three-dimensional active matter.

The remainder of the paper is organized as follows. In Sec.~II we classify isotropic odd currents and derive the unique lowest-order nonlinear constitutive law allowed in three dimensions. In Sec.~III we analyze its consequences, emphasizing the emergence of boundary-driven rotational transport, vorticity, and enstrophy. In Sec.~IV we derive the theory from a microscopic model with nonreciprocal interactions. Technical derivations and alternative geometric formulations are in the Appendices.

\section{Symmetry Construction of Three-Dimensional Odd Diffusion}

The existence of odd transport in isotropic media is strongly constrained by symmetry. In two dimensions, parity-odd transport coefficients arise naturally because rotational invariance permits the antisymmetric tensor $\epsilon_{ij}$. This tensor allows density gradients to be rotated by ninety degrees, leading to transverse currents that do not derive from a free-energy functional. Odd viscosity, odd elasticity, odd mobility, and odd diffusion all exploit this geometric structure. The situation in three dimensions is fundamentally different. While the three-dimensional Levi--Civita tensor $\epsilon_{ijk}$ exists, it carries three indices rather than two. Consequently, no isotropic antisymmetric rank-two tensor exists from which a linear odd diffusion law can be constructed. This observation appears to prohibit isotropic odd diffusion in three dimensions.

In this section, we show that this conclusion is not fundamental. The obstruction applies only to linear constitutive laws. By considering nonlinear couplings between multiple conserved fields, rotational invariance permits a unique parity-odd transport term that remains fully isotropic in three dimensions.

\subsection{Failure of Linear Odd Diffusion in Three Dimensions}

Consider first a single conserved scalar density $\rho(\mathbf r,t)$ whose flux is assumed to depend linearly on gradients,
\begin{equation}
J_i=-M_{ij}\partial_j \rho 
\label{eq:linearcurrent}
\end{equation}
where $M_{ij}$ is a transport matrix that can be decomposed into symmetric and antisymmetric parts,
\begin{equation}
M_{ij}=D_{ij}+A_{ij},
\end{equation}
where
\begin{equation}
D_{ij}=D_{ji},
\qquad
A_{ij}=-A_{ji}.
\end{equation}
For an isotropic medium in three dimensions, rotational invariance requires that every rank-two tensor be proportional to the identity \cite{jeffreys1973isotropic,epstein2020time},
\begin{equation}
D_{ij}=D \delta_{ij},
\end{equation}
and thus no nonzero isotropic antisymmetric rank-2 tensor exists in three dimensions. Consequently,
\begin{equation}
A_{ij}=0,
\end{equation}
and the only allowed linear constitutive law is ordinary diffusion,
\begin{equation}
J_i=-D\partial_i\rho .
\end{equation}
This argument establishes that isotropic linear odd diffusion is forbidden in three dimensions. The absence of linear odd diffusion does not imply the absence of odd transport altogether. Rather, it motivates the search for nonlinear constitutive relations that circumvent the rank-two tensor obstruction.

\subsection{Minimal Nonlinear Construction}

To identify the lowest-order odd contribution, we consider a system containing multiple conserved densities $\rho^A(\mathbf r,t)$, where the species index $A$ labels distinct conserved components. The desired current must satisfy four requirements: rotational invariance, conservation of each species, parity oddness, and locality in the density fields.  The only isotropic pseudotensor available in three dimensions is $\epsilon_{ijk}$. To construct a vector current, two additional vectors are therefore required. The lowest-order choice is provided by gradients of the densities, $\partial_i\rho^A$ . Combining these objects yields the parity-odd current
\begin{equation}
J_{i,\mathrm{odd}}^A= -\omega \epsilon_{ijk} \epsilon^{ABC} (\partial_j\rho^B)(\partial_k\rho^C),
\label{eq:oddcurrent}
\end{equation}
where $\epsilon^{ABC}$ is the antisymmetric rank-3 tensor in species space, and $\omega$ is the odd diffusion parameter, which we assume to be a constant and is independent of species. Adding this flux to the ordinary diffusive contribution gives
\begin{equation}
J_i^A=-D\partial_i\rho^A -\omega \epsilon_{ijk} \epsilon^{ABC} (\partial_j\rho^B) (\partial_k\rho^C),
\label{eq:constitutive}
\end{equation}
which defines a non-linear, odd diffusive constitutive relation. Equation \ref{eq:constitutive} is the central result of this work.

Unlike conventional odd diffusion, the current is nonlinear in density gradients and requires at least three conserved fields. 
The structure of Eq.~\ref{eq:constitutive} is not arbitrary. At lowest order in gradients, rotational invariance requires every parity-odd contribution to contain exactly one Levi--Civita tensor. Conservation demands that the current carries a single free spatial index, while isotropy forbids the introduction of fixed vectors or tensors. These constraints uniquely select Eq.~\ref{eq:oddcurrent} as the lowest-order isotropic odd contribution. Any alternative construction either vanishes identically, reduces to a total derivative, violates isotropy, or appears only at higher order in gradients. A complete tensor classification is presented in Appendix \ref{App:A}.

\begin{figure}
    \centering
    \includegraphics[width=1.\linewidth]{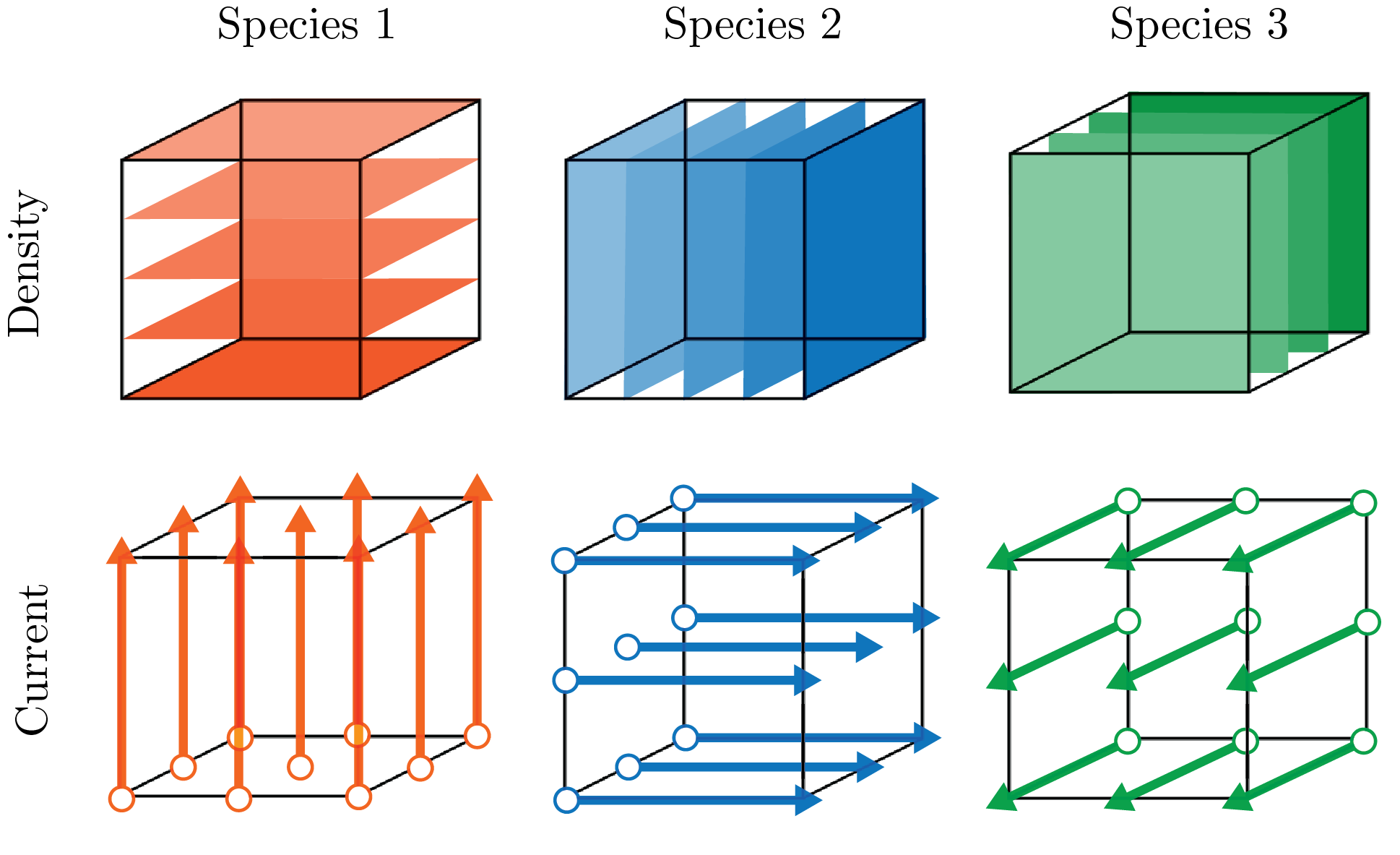}
    \caption{Illustration of odd diffusive currents. For a set of density fields shown in the top row for which opacity reflects the change to the density, the odd contribution of the current is illustrated in the bottom row. For $\omega>0$, each species exhibits a current that is perpendicular to the plane given by the direction of the density gradient of the other two species.}
      \label{fig:schematic}
\end{figure}

The odd contribution to the flux can be written compactly as a cross product,
\begin{equation}
\mathbf J^A_{\mathrm{odd}}=-\omega \epsilon^{ABC} \nabla\rho^B \times \nabla\rho^C 
\label{eq:crossform}
\end{equation}
which leads itself to several different geometric interpretations. First, the current in species $A$ is perpendicular to density gradients in species $B$ and $C$. Consequently, it transports material transverse to the thermodynamic forces driving diffusion. In this respect, it resembles Hall transport and conventional odd diffusion. Second, the current is intrinsically three-dimensional. The cross product requires two independent gradients whose mutual orientation determines the direction of transport. This directionality is illustrated in Fig.~\ref{fig:schematic}. The resulting flow therefore probes geometric information that is unavailable in two-dimensional odd diffusion.
Finally, Eq.~\ref{eq:crossform} implies that odd transport is generated only when gradients of distinct species fail to remain parallel. Regions where concentration gradients align produce no odd current; whereas, for those regions are nonparallel, they generate rotational transport. This geometric structure underlies the unusual dynamical properties discussed in the following section on boundary driven transport.

\section{Boundary-Driven Rotational Transport}
The constitutive law introduced in the previous section has a divergence that vanishes identically in the bulk. Consequently, the odd contribution does not alter the steady-state density profiles of the conserved fields. Instead, it generates rotational currents whose structure is determined entirely by boundary conditions. This property leads to a distinctive form of nonequilibrium transport. The densities obey the same harmonic equations as ordinary diffusion, yet the system supports finite circulation, vorticity, and enstrophy. Odd transport therefore survives not through modification of the density field, but through the geometry of the associated current patterns.

\subsection{Divergence-Free Structure}

A defining property of the odd current is revealed by rewriting Eq.~\ref{eq:constitutive} in vector notation, as in Eq. \ref{eq:crossform}. The divergence of this current follows immediately from a standard vector identity,
$\nabla\cdot \left( \nabla f \times \nabla g \right) = 0$, valid for arbitrary smooth scalar fields $f$ and $g$ \cite{tong2025electromagnetism}. Substituting Eq.~\ref{eq:crossform} yields
\begin{equation}
\nabla\cdot \mathbf J_{\mathrm{odd}}^A=0
\label{eq:divfree}
\end{equation}
which does not rely on any approximation or special choice of density profile and is exact assuming that the odd parameter $\omega$ is a constant. Therefore, the odd contribution generates transport without producing local sources or sinks. Physically, Eq.~\ref{eq:divfree} implies that the odd current redistributes material only through circulation. Material transported in one direction must necessarily return elsewhere, producing closed current loops. The current can therefore be substantial even though it makes no direct contribution to local density relaxation. The odd current survives as a purely solenoidal component of the flux. As a consequence, the conservation law for each species takes the form
\begin{equation}
\partial_t \rho^A=-\nabla\cdot \mathbf J^A = D\nabla^2\rho^A
\label{eq:laplace}
\end{equation}
as in normal linear diffusion. In steady state, therefore, each density satisfies the Laplace equation and is identical to that obtained for ordinary diffusion.

\begin{figure*}
    \centering
    \includegraphics[width=1.\linewidth]{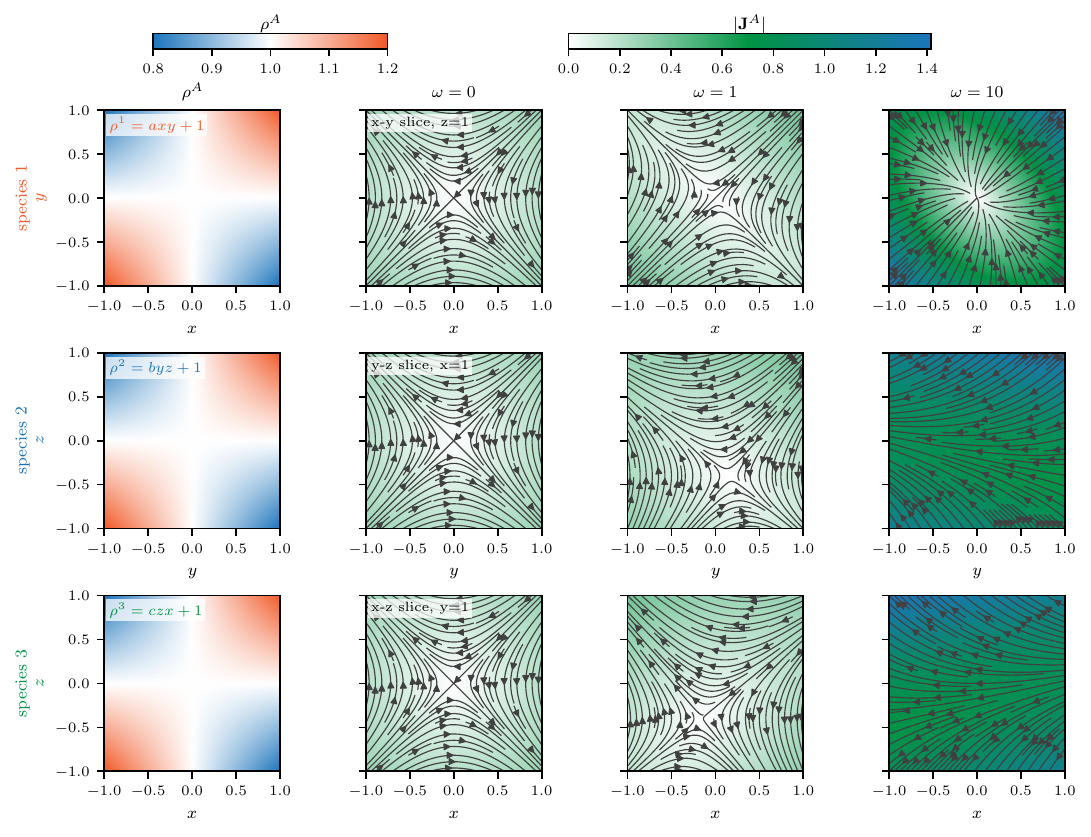}
    \caption{Odd diffusion from multiple species in a system with linear size $2L$. Left panels display fixed density profiles consistent with Laplace's equation for $a=b=c=1/5$. Right panels illustrate the resultant currents computed from the constitutive relationship with $L=D=1$ and density profiles in left panels. Left to right show the results for increasing odd diffusion coefficient, $\omega$. Each row contains information for a specific species.  Densities are dimensionless. 
 }
\label{fig:currentstreamlinesnew}
\end{figure*}
Equation~\ref{eq:laplace} might suggest that odd transport is physically irrelevant. Such a conclusion would be incorrect. While the densities are unchanged, the odd behavior survives as a finite rotational flow superimposed on the harmonic density field. Consequently, two systems with identical concentration profiles may exhibit entirely different transport patterns depending on the magnitude of the odd coefficient $\omega$. This separation between density and current constitutes one of the central features of three-dimensional odd diffusion. The observable consequences of odd transport appear not in the concentration field itself, but in the geometry of the associated fluxes. Moreover, while odd diffusion necessitates breaking the time reversal symmetry for $\omega\neq 0$, because $\mathbf{J}_{\mathrm{odd}}^A$ is divergence free, odd diffusion does not contribute to entropy production in a bulk fluid directly.

\subsection{Boundary-Driven Currents}

Because the densities satisfy Laplace's equation, all information about the odd transport is encoded through the boundary conditions. This observation implies that boundaries play a fundamentally different role than in conventional diffusion. In ordinary diffusion, boundaries determine the concentration field, which in turn determines transport. Here, boundaries determine both the concentration field and the topology of the resulting current loops. The odd current therefore represents a nonlocal geometric response to the arrangement of boundary concentrations.

\begin{figure*}
    \centering
    \includegraphics[width=1.\linewidth]{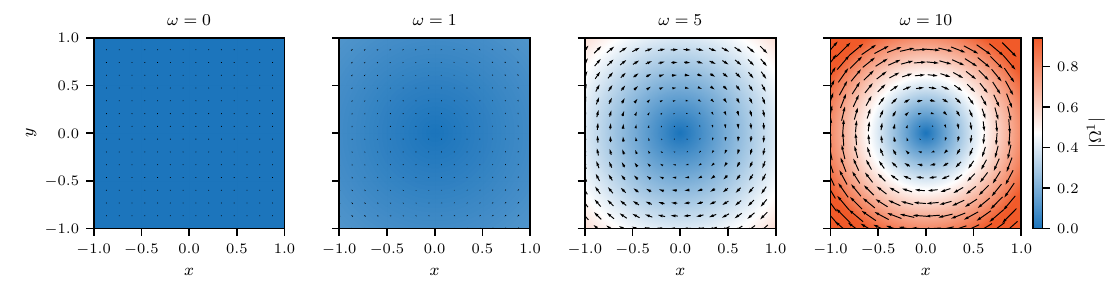}
    \caption{Vorticity generation from odd diffusion. Vorticity of species 1 for increasing odd diffusion coefficient $\omega$, computed from the density profiles in Fig.~\ref{fig:currentstreamlinesnew}. Colorbar highlights the magnitude of the vorticity in the $x-y$ plane for fixed $z=L=1$, while arrows denote its direction. 
    }
    \label{fig:currentandvorticity}
\end{figure*}
To illustrate this mechanism, consider three species whose boundary values generate approximately orthogonal concentration gradients throughout the domain. The harmonic solutions remain simple linear functions of position. Nevertheless, Eq.~\ref{eq:crossform} predicts finite circulating currents whose direction is fixed by the handedness of the gradient triad. Reversing the ordering of the species reverses the direction of circulation, reflecting the parity-odd nature of the transport coefficient. The resulting flow pattern resembles a stationary vortex despite the absence of momentum conservation, fluid advection, or externally imposed torque. Rotational transport therefore emerges purely from the geometry of concentration gradients.

Figure~\ref{fig:currentstreamlinesnew} illustrates representative examples of these boundary-induced current loops. In all cases, the density fields remain harmonic while the odd current forms nontrivial circulation patterns throughout the bulk. These examples demonstrate that the principal observable consequence of three-dimensional odd diffusion is not altered density relaxation but rather the emergence of boundary-controlled rotational transport.

\subsection{Vorticity Generation}
The divergence-free structure of the odd current implies that it transports material through circulation rather than accumulation. A natural quantity characterizing such circulation is the vorticity $\boldsymbol{\Omega}^{A}$ of the current,
\begin{equation}
\boldsymbol{\Omega}^A = \nabla\times \mathbf J_{\mathrm{odd}}^A 
\label{eq:vorticitydef}
\end{equation}
and is computable from the curl of the odd current. Consequently, the transport generated by Eq.~\ref{eq:crossform} exhibits a rotational character even in the absence of momentum conservation, fluid advection, or externally imposed torques. Substituting the constitutive law into Eq.~\ref{eq:vorticitydef} for steady-state density profiles satisfying Laplace's equation
\begin{align}
\boldsymbol{\Omega}^A =& -\omega \epsilon^{ABC} \left(\mathbf{H}^B \nabla\rho^C-\mathbf{H}^C \nabla\rho^B \right) \, ,
\label{eq:vorticitygeneral}
\end{align}
where 
\begin{equation}
H^A_{ij} =\partial_i\partial_j\rho^A \, .
\end{equation}
Here, $\mathbf{H}^A$ denotes the Hessian matrix of $\rho^A$. Equation~\ref{eq:vorticitygeneral}  reveals that vorticity originates from spatial variation in the orientation of the concentration gradients. Regions where all gradients remain uniform generate circulating currents but no local vorticity. By contrast, curved harmonic profiles produce finite rotational structure throughout the bulk. The vorticity is controlled by the competition between local curvature of one concentration field and gradients of another. 

This observation highlights an important distinction between the present mechanism and conventional hydrodynamic vortices. In ordinary fluids, vorticity is generated through momentum transport, external forcing, or boundary shear \cite{landau1987fluid}. Here, vorticity is generated entirely through the geometry of concentration fields. The system need not possess inertia or conserve momentum. The rotational structure is instead encoded directly in the constitutive law.

To illustrate this mechanism, consider three species whose boundary conditions generate nonlinear harmonic profiles. Although each density independently satisfies Laplace's equation, the gradients are generally nonuniform and intersect at varying angles throughout the domain. From those concentration profiles, Eq.~\ref{eq:vorticitygeneral} predicts finite vorticity concentrated in regions where the gradients exhibit the strongest curvature. Reversing the handedness of the concentration pattern changes the sign of the vorticity, reflecting the parity-odd nature of the transport coefficient. The resulting current field resembles a stationary network of vortices whose orientation is determined entirely by the boundary concentrations. These structures are not transient features associated with relaxation dynamics, but persistent signatures of nonlinear odd diffusion.

Representative examples of the vorticity field are shown in Fig.~\ref{fig:currentandvorticity}. These calculations demonstrate that substantial rotational structure can emerge even when the underlying concentration profiles remain simple harmonic functions. The existence of such boundary-controlled vortices constitutes one of the most distinctive consequences of isotropic odd diffusion in three dimensions.

\subsection{Global Measures of Odd Transport}
The vorticity field introduced above provides a local characterization of rotational transport. To compare different geometries, boundary conditions, and transport coefficients, it is useful to introduce a global measure of the strength of odd transport. A natural choice is the enstrophy $\mathcal E$ \cite{doering1995applied}, defined as
\begin{equation}
\mathcal E = \frac{1}{2} \sum_A \int_V d \mathbf{r}  \left| \boldsymbol{\Omega}^A \right|^2 ,
\label{eq:enstrophy}
\end{equation}
the normal of the vorticity integrated over the system volume $V$. In conventional fluid mechanics, enstrophy measures the intensity of vortical motion and plays an important role in the characterization of turbulent and coherent structures \cite{jimenez2018coherent,buaria2020vortex}. Here, Eq.~\ref{eq:enstrophy} serves a different purpose. Because the density fields remain harmonic, the enstrophy quantifies the rotational component of transport that is invisible at the level of concentration profiles. It therefore provides a direct measure of the nonequilibrium consequences of the odd constitutive law.

Substituting Eq.~\ref{eq:vorticitygeneral} into Eq.~\ref{eq:enstrophy} gives
\begin{equation}
\mathcal E = \frac{\omega^2}{2} \sum_A \int_V d\mathbf{r} \left| \epsilon^{ABC} \left (\mathbf{H}^B\nabla\rho^C -  \mathbf{H}^C\nabla\rho^B\right ) \right|^2  
\label{eq:enstrophygeneral}
\end{equation}
from which several immediate consequences follow. First, the enstrophy scales quadratically with the odd transport coefficient, $\mathcal E \propto \omega^2$, consequently, reversing the sign of $\omega$ changes the handedness of the current loops but leaves the total magnitude of rotational transport unchanged. Second, Eq.~\ref{eq:enstrophygeneral} vanishes whenever all concentration gradients are spatially uniform. In this case the current may remain finite, but the associated transport is irrotational. Enstrophy therefore isolates the genuinely vortical component of the odd response.  Third, the enstrophy is controlled entirely by the geometry of the harmonic density fields. Since these fields are themselves determined by the boundary conditions, Eq.~\ref{eq:enstrophygeneral} establishes a direct connection between boundary concentrations and the global strength of rotational transport. In this sense, enstrophy acts as a compact scalar descriptor of the boundary-induced current topology. An especially revealing feature of Eq.~\ref{eq:enstrophygeneral} is that it remains finite even though the density evolution obeys the same diffusion equation as an ordinary passive mixture. Two systems may therefore possess identical steady-state concentration profiles while exhibiting dramatically different values of $\mathcal E$. From the perspective of the density field alone, the systems are indistinguishable, yet their transport properties are fundamentally different.

The existence of finite enstrophy in a system whose densities satisfy Laplace's equation underscores the unusual nature of isotropic odd diffusion in three dimensions. While ordinary diffusion relaxes concentration gradients without generating rotational transport, the nonlinear odd current converts the geometric arrangement of those gradients into persistent circulating flows. Enstrophy provides a quantitative measure of that conversion and furnishes a natural observable for experiments and simulations seeking to identify three-dimensional odd transport. Specific calculations of the vorticity and enstrophy are shown in Appendix \ref{App:B}.

\section{Microscopic Origin}

The constitutive law introduced in Sec.~II was derived from symmetry considerations alone. While symmetry identifies the form of transport permitted at long wavelengths, it does not establish whether such behavior can emerge from a microscopic dynamical system. In this section, we demonstrate that the nonlinear odd current arises naturally from a simple class of isotropic nonreciprocal interactions between multiple particle species. Our goal is not to identify a unique microscopic realization. Rather, we seek the minimal microscopic ingredients required to generate isotropic odd transport in three dimensions. As we show below, nonreciprocal three-body interactions provide precisely such a mechanism.

\subsection{Nonreciprocal Three-Body Interaction Model}

\begin{figure}
    \centering
    \includegraphics[width=1\linewidth]{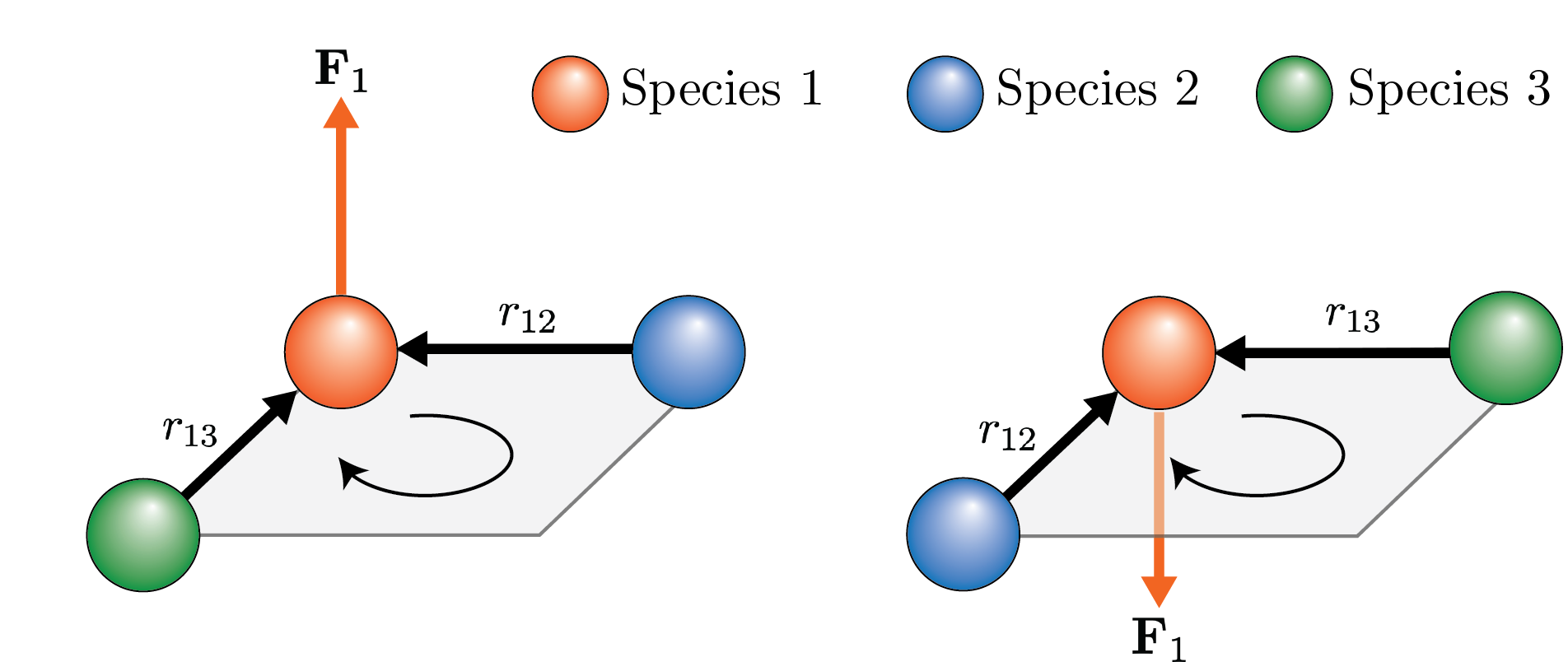}
    \caption{Schematic representation of the chiral, nonreciprocal three-body force, Eq. \ref{eq:threebodyforce} from a single interacting triple, with one
particle drawn from each species: particle $1\in A$ (red), particle
$2\in B$ (green), and particle $3\in C$ (blue). (left) The force on particle~$1$,
$\mathbf{F}_1 \propto \epsilon^{ABC}\,\mathbf{r}_{1,2}\times\mathbf{r}_{1,3}$,
is the cross product of the two displacement vectors
$\mathbf{r}_{1,2}=\mathbf{r}_1-\mathbf{r}_2$ and
$\mathbf{r}_{1,3}=\mathbf{r}_1-\mathbf{r}_3$. As the cross product of two in-plane vectors, $\mathbf{F}_1$
(red) is directed out of the plane of the triple; the curved arrow marks the
cyclic species order $A\!\to\!B\!\to\!C$ that fixes its sign. (right)
Exchanging the two species $B$ and $C$ (cyclic order $A\!\to\!C\!\to\!B$)
reverses $\epsilon^{ABC}\!\to\!-\epsilon^{ABC}$, and $\mathbf{F}_1$ flips to
point in the opposite direction.}
    \label{fig.threebodyforce}
  
\end{figure}

Consider a collection of overdamped particles belonging to three conserved species labeled by $A\in{1,2,3}$. The position of particle $\alpha$ of type $A$ evolves according to
\begin{equation}
\dot{\mathbf r}_{\alpha} = \mu \mathbf F_{\alpha} + \sqrt{2D} \boldsymbol{\eta}_{\alpha} ,
\label{eq:langevin}
\end{equation}
where $\mu$ is the mobility, $D$ is the diffusivity, and $\boldsymbol{\eta}_\alpha$ is Gaussian white noise with $\langle \eta_{\alpha,i}(t)\rangle =0$ and
\begin{equation}
\langle \eta_{\alpha,i}(t) \eta_{\beta,j}(t') \rangle =\delta_{\alpha \beta} \delta_{ij} \delta(t-t')
\end{equation}
independent and time local. 
Ordinary equilibrium systems derive their interactions from pair potentials and therefore satisfy Newton's third law. Such interactions cannot generate the parity-odd transport identified in Sec.~II. We instead consider nonreciprocal interactions involving triples of particles belonging to distinct species. Species are encoded through projectors
\begin{equation}
S_\alpha^A=\delta_{A,\sigma(\alpha)},
\end{equation}
which satisfy
\begin{equation}
S_\alpha^A S_\alpha^B=\delta_{AB}S_\alpha^A, \qquad \sum_A S_\alpha^A =1.
\end{equation}
Assuming the particles interact through a nonreciprocal microscopic force, the lowest-order isotropic force compatible with parity breaking on particle $\alpha$ is
\begin{equation}
\mathbf{F}_{\alpha}= \lambda  \varepsilon^{ABC} \sum_{\beta,\gamma} S_\alpha^A S_\beta^B
S_\gamma^C G_{\alpha\beta\gamma}  \mathbf{r}_{\alpha,\beta} \times  \mathbf{r}_{\alpha,\gamma} ,
%
\label{eq:threebodyforce}
\end{equation}
where $G_{\alpha\beta\gamma}( | \mathbf r_{\alpha, \beta} |,  | \mathbf r_{\alpha,\gamma} |)$ is a scalar three-body interaction kernel that depends on the norm of the displacement $\mathbf r_{\alpha,\beta}= \mathbf r_{\alpha}- \mathbf r_{\beta}$ between particles $\alpha$ and $\beta$ that acts at short range, and $\lambda$ controls the strength of the nonreciprocal coupling. The Levi--Civita tensor in species space ensures that the interaction changes sign under exchange of species labels and vanishes when any two species coincide. A schematic of the three-body force is shown in Fig. \ref{fig.threebodyforce}.

Unlike equilibrium pair interactions, Eq.~\ref{eq:threebodyforce} is not derivable from a potential. Detailed balance is therefore broken at the microscopic level, providing the nonequilibrium ingredient required for odd transport, and a corresponding breaking of time-reversal symmetry \cite{limmer2024statistical}. The interaction is nevertheless isotropic in physical space. No preferred direction is introduced, and any parity breaking arises dynamically through the coupling between species.

\subsection{Dean--Kawasaki coarse-graining}
To connect the microscopic dynamics to the continuum theory, we introduce the empirical density fields
\begin{equation}
\hat{\rho}^A(\mathbf r,t) = \sum_{\alpha\in A} \delta [ \mathbf r-\mathbf r_\alpha(t) ],
\label{eq:densityfield}
\end{equation}
which fluctuates in time due to the evolution of the particles. Applying standard Dean--Kawasaki methods \cite{dean1996langevin} yields an exact stochastic evolution equation for the density fields,
\begin{equation}
\partial_t\hat{\rho}^A(\mathbf r,t) = -\nabla\cdot  \hat{\mathbf{J}}^A(\mathbf r,t) +\nabla \cdot \sqrt{2 D } \hat{\rho}^A(\mathbf r,t)\xi(\mathbf r,t),
\label{eq:DK}
\end{equation}
where $\xi(\mathbf r,t)$ denotes a multiplicative noise field, with correlations  
\begin{equation}
\langle \xi_i(\mathbf r,t)\xi_j(\mathbf r',t') \rangle =\delta_{ij}\delta(\mathbf{r}-\mathbf{r}') \delta(t-t').
\end{equation}
and the fluctuating current, $\hat{J}^A(\mathbf r,t)$, contains contributions from diffusion and interactions.

The full derivation is lengthy and is presented in Appendix~\ref{App:C}. Here we emphasize only the essential physical mechanism. The three-body interaction generates terms involving products of density gradients belonging to distinct species. After a gradient expansion and closure of higher-order correlation functions, the leading parity-odd contribution to the current becomes
\begin{equation}
\mathbf J_{\mathrm{odd}}^A =- \mu \lambda \rho^A \mathcal I \epsilon^{ABC} \nabla\rho^B \times \nabla\rho^C ,
\label{eq:derivedodd}
\end{equation}
where $\mathcal I$ is a moment of the microscopic interaction kernel
\begin{equation}
\mathcal I= \frac{1}{6} \int d\mathbf{r} d\mathbf{r}' \, G(\mathbf r,\mathbf r') |\mathbf r\times\mathbf r'|^2 
\label{eq:omegafinal}
\end{equation}
determined by the initial non-reciprocal force. To make contact with the constitutive law Eq.~\ref{eq:maincurrent}, we expand each density around a uniform background $\rho^A(\mathbf{r}, t)=\bar{\rho}+\delta \rho^A(\mathbf{r}, t)$. Since the odd current is quadratic in density fields, to leading order, we obtain 
\begin{equation}
\mathbf J_{\mathrm{odd}}^A =- \mu \lambda \bar\rho \mathcal I \epsilon^{ABC} \nabla\rho^B \times \nabla\rho^C ,
\end{equation}
which is identical to the constitutive law obtained from symmetry in Sec.~II, when we identify $\omega = \mu \lambda \mathcal I \bar\rho$.

\subsection{Emergent Odd Transport}

The derivation above establishes that the nonlinear odd current is not an abstract symmetry construction but rather the generic hydrodynamic signature of isotropic nonreciprocal three-body interactions. The microscopic origin of the transport can be understood intuitively.
The interaction kernel couples local gradients of distinct species and generates a tendency for particles to move perpendicular to the plane defined by neighboring concentration differences. When coarse grained over many particles, this geometric bias accumulates into the cross-product structure appearing in Eq.~\ref{eq:derivedodd}. The resulting transport remains isotropic because no preferred spatial direction exists. Instead, the direction of the current is determined dynamically by the local arrangement of concentration gradients.

The derivation also clarifies why multiple conserved fields are required. The cross-product structure arises from interactions between gradients of distinct species. A single-component system lacks the geometric information needed to construct an isotropic parity-odd vector current. The existence of three-dimensional odd transport is therefore inseparable from the multicomponent nature of the underlying dynamics.
Taken together, these results provide a microscopic foundation for the hydrodynamic theory developed above. The constitutive law follows simultaneously from symmetry and from coarse-graining of a minimal nonequilibrium particle model. The emergence of boundary-controlled vorticity and enstrophy can therefore be viewed as a robust macroscopic consequence of isotropic nonreciprocal interactions rather than a peculiarity of a particular continuum construction.

\section{Discussion and Conclusions}

Odd transport coefficients have emerged as a defining feature of nonequilibrium chiral matter. Existing examples are most naturally formulated in two dimensions, where the antisymmetric tensor $\epsilon_{ij}$ provides a simple mechanism for generating currents transverse to thermodynamic forces. This geometric structure has led to the widespread view that isotropic odd diffusion is fundamentally a two-dimensional phenomenon.
The results presented here demonstrate that this conclusion is not general. While isotropic \emph{linear} odd diffusion is indeed forbidden in three dimensions, nonlinear couplings between multiple conserved fields circumvent the underlying symmetry obstruction. By exploiting the three-dimensional Levi--Civita tensor together with gradients of distinct species, we identified the unique lowest-order isotropic parity-odd current permitted by rotational invariance. This constitutive law establishes a minimal framework for odd transport in isotropic three-dimensional systems.

The resulting phenomenology differs qualitatively from conventional odd diffusion. The odd current is divergence-free and therefore does not contribute directly to the bulk evolution of the conserved densities. As a consequence, the density fields obey the same harmonic equations as ordinary diffusion and cannot, by themselves, reveal the presence of odd transport. Instead, the nonequilibrium character of the dynamics appears through the geometry of the current field. Harmonic concentration profiles generate persistent circulating currents, finite vorticity, and nonzero enstrophy even in the absence of externally imposed torques or preferred spatial directions. This observation suggests a distinct perspective on odd transport in three dimensions. Rather than modifying density relaxation, the odd contribution reorganizes transport into rotational structures whose existence is determined entirely by boundary conditions. The boundaries prescribe harmonic concentration fields, while the nonlinear constitutive law converts the relative orientation of those fields into circulating currents. 

From a theoretical standpoint, the present construction reveals a broader principle. Symmetry restrictions that prohibit linear nonequilibrium responses do not necessarily prohibit nonlinear responses. Nonlinear transport relations generally allow more flexible form since their are not constrained by thermodynamics \cite{gao2019nonlinear,limmer2021large,deshpande2024odd}. In the present case, the absence of an isotropic antisymmetric rank-two tensor forbids conventional odd diffusion but leaves open a nonlinear route through multicomponent couplings. Similar mechanisms may arise in other classes of active and driven systems where symmetry appears to exclude parity-odd transport at linear order. More generally, our results suggest that the classification of odd hydrodynamic responses may be substantially richer than is apparent from linear constitutive theories alone. A simple class of isotropic nonreciprocal three-body interactions generates precisely the constitutive law predicted by symmetry after coarse-graining. This establishes a direct connection between microscopic violations of detailed balance and macroscopic rotational transport. The mechanism requires neither external fields nor intrinsic particle rotation and therefore provides a route toward realizing odd transport in systems that remain isotropic at large scales. Moreover, the microscopically generated current also resembles the structure of skyrmion topological density which points towards the potential connection in topological field theory \cite{neubauer2009topological}. Candidate systems include multicomponent active colloids, synthetic reaction--diffusion media with nonreciprocal interactions, driven chemical networks, and biological systems containing multiple interacting conserved species \cite{tucci2024nonreciprocal,bililign2022motile,mecke2023simultaneous}.

Several extensions of the present theory deserve further investigation. The analysis presented here focused on deterministic hydrodynamic transport and steady-state behavior. Incorporating fluctuations into the continuum description may reveal novel nonequilibrium correlations associated with the divergence-free odd current. Likewise, coupling the concentration fields to momentum-conserving hydrodynamics could generate new classes of active vortical states in which odd diffusion and fluid flow interact self-consistently. The odd current is naturally expressed as the cross product of concentration gradients and therefore depends on the spatial arrangement of multiple conserved fields. This raises the possibility of topologically protected transport structures, linked current loops, and other geometric phenomena that have no analogue in conventional diffusion. The geometric formulation developed in the Appendices suggests a natural language for exploring such questions.


\appendix

\section{Symmetry Classification of Isotropic Odd Transport}\label{App:A}

In this Appendix we derive the most general isotropic parity-odd contribution to the current at lowest order in gradients and establish the uniqueness of Eq.~\ref{eq:constitutive} at lowest order in the gradient expansion. Consider a conserved scalar field $\rho$ whose current depends linearly on gradients,
\begin{equation}
J_i=-M_{ij}\partial_j\rho.
\end{equation}
where the rank two tensor $M_{ij}$ can be decomposed as
\begin{equation}
M_{ij}=D_{ij}+A_{ij},
\end{equation}
where
\begin{equation}
D_{ij}=D_{ji},
\qquad
A_{ij}=-A_{ji}.
\end{equation}
For an isotropic medium, in three dimensions, rotational invariance requires
\begin{equation}
D_{ij}=D\delta_{ij}.
\end{equation}
The only rotationally invariant rank-two tensor is therefore the identity. Since no isotropic antisymmetric rank-two tensor exists in three dimensions,
\begin{equation}
A_{ij}=0.
\end{equation}
Consequently, the most general isotropic linear constitutive law is
\begin{equation}
\mathbf J=-D\nabla\rho.
\end{equation}
This proves that isotropic linear odd diffusion is forbidden in three dimensions.

We now seek parity-odd currents constructed from gradients of multiple conserved densities $\rho^A$. The current must 1) carry one free spatial index, 2) transform as a vector under rotations, 3) change sign under parity,
and 4) vanish when parity symmetry is restored. The only isotropic pseudotensor available is $\epsilon_{ijk}$. Therefore every parity-odd current must contain a spatial Levi--Civita tensor.

At lowest order in gradients the only possible construction is
\begin{equation}
J_i^A=- C^{ABC} \epsilon_{ijk} (\partial_j\rho^B) (\partial_k\rho^C).
\end{equation}
Because $\epsilon_{ijk}$ is antisymmetric under interchange of $j$ and $k$, only the antisymmetric component of $C^{ABC}$ contributes,
\begin{equation}
C^{ABC} = -C^{ACB}.
\end{equation}
For three species the unique isotropic antisymmetric tensor is
\begin{equation}
C^{ABC}=\omega \epsilon^{ABC},
\end{equation}
where $\omega$ is a constant, which yields Eq. \ref{eq:oddcurrent} up to an overall coefficient, $\omega$, this is the unique parity-odd contribution appearing at second order in gradients.

Additional isotropic parity-odd currents may be constructed at higher order by introducing additional gradients or Laplacians. Examples include
\begin{equation}
\epsilon_{ijk} (\nabla^2\partial_j\rho^B) (\partial_k\rho^C),
\end{equation}
and
\begin{equation}
\epsilon_{ijk} (\partial_j\rho^B) (\partial_k\nabla^2\rho^C).
\end{equation} 
These contributions are sub-leading in the long-wavelength limit and therefore do not modify the leading constitutive law discussed in the main text.

\section{Explicit Vorticity and Enstrophy Calculations}\label{App:B}

This Appendix presents representative examples illustrating the emergence of vorticity and enstrophy from harmonic density fields.

\subsection{Uniform Gradient Profiles}

Consider boundary conditions that prescribe the following linear density profiles
\begin{align}
\rho^1 &= a x,\\
\rho^2 &= b y,\\
\rho^3 &= c z
\end{align}
where $a,b,$ and $c$ are constants proportional to the inverse of the linear domain size. Correspondingly, the gradients are constant,
\begin{align}
\nabla\rho^1 &= a\hat x,\\
\nabla\rho^2 &= b\hat y,\\\
\nabla\rho^3 &= c\hat z.
\end{align}
which generates an odd current
\begin{equation}
\mathbf J_{\rm odd}^1=-2\omega bc \hat x,
\end{equation}
with analogous expressions for the remaining species.

Although the current is finite, its curl vanishes,
\begin{equation}
\nabla\times\mathbf J_{\rm odd}^A=0 
\end{equation}
and consequently,
\begin{equation}
\mathcal E=0
\end{equation}
the enstrophy vanishes. This example demonstrates that odd transport need not imply vorticity.

\subsection{Quadratic Harmonic Profiles}

Consider  the following harmonic, nonlinear density profiles
\begin{align}
\rho^1 &= axy,\\
\rho^2 &= byz,\\
\rho^3 &= czx
\end{align}
each of which satisfies Laplace's equation,
\begin{equation}
\nabla^2\rho^A=0.
\end{equation}
and thus are stationary.  The gradients become
\begin{align}
\nabla\rho^1 &=(ay,ax,0),\\
\nabla\rho^2 &=(0,bz,by),\\
\nabla\rho^3 &=(cz,0,cx)
\end{align}
which implies an odd current for species $1$,
\begin{equation}
\mathbf J_{\rm odd}^1=-2bc \omega (zx,yz,-z^2)
\end{equation}
and the associated vorticity is
\begin{equation}
\label{eq:omegaHP}
\boldsymbol{\Omega}^1 = 2 b c \omega (y,-x,0).
\end{equation}
Unlike the uniform-gradient example, the vorticity is nonzero throughout the bulk.

For a cubic domain of side length $L$ anchored at the origin, the enstrophy can be evaluated by substituting Eq.~\ref{eq:omegaHP} into its definition and integrating
\begin{equation}
\mathcal E = \frac{4\omega^2 L^5}{3} \left (a^2 b^2 +a^2 c^2 + b^2 c^2 \right )
\end{equation}
which shows explicitly that the enstrophy is finite and scales quadratically with the odd transport coefficient. For coefficients $a,b$ and $c$ proportional to $1/L$, the enstrophy  scales sub-extensively, as the linear extent of the system.

\section{Dean--Kawasaki Derivation of the Odd Constitutive Law}\label{App:C}
In this Appendix, we derive the nonlinear odd current appearing in Eq.~\ref{eq:constitutive} from a microscopic model of interacting Brownian particles. The derivation establishes an explicit connection between isotropic nonreciprocal three-body interactions and the macroscopic odd transport coefficient.

\subsection{Microscopic Dynamics}
We consider a collection of overdamped Brownian particles belonging to three species labeled by $A,B,C\in{\{1,2,3\}}$ in three spatial dimensions $\mathbf{r}=\{r_i,r_j,r_k\}$. Particle labels are denoted by Greek indices $\alpha,\beta,\gamma$. The overdamped Langevin dynamics are \cite{limmer2024statistical}
\begin{equation}
\dot{r}_{\alpha,i}= \mu F_{\alpha,i}+\sqrt{2D}\eta_{\alpha,i}(t),
\label{eq:app_langevin}
\end{equation}
with $\langle \eta_{\alpha,i}(t)\rangle =0$ and
\begin{equation}
\left< \eta_{\alpha,i}(t) \eta_{\beta,j}(t') \right>=\delta_{\alpha\beta} \delta_{ij} \delta(t-t').
\end{equation}
noise that is white and uncorrelated. Species are encoded through projectors
\begin{equation}
S_\alpha^A=\delta^A_{\sigma(\alpha)},
\end{equation}
which satisfy
\begin{equation}
S_\alpha^A S_\alpha^B=\delta^{AB}S_\alpha^A, \qquad \sum_A S_\alpha^A =1.
\end{equation}
As in the main text, we assume the particles interact through a nonreciprocal microscopic force. For concreteness we take the force on particle $\alpha$ to be
\begin{equation}
\mathbf{F}_{\alpha}= \lambda  \varepsilon^{ABC} \sum_{\beta,\gamma} S_\alpha^A S_\beta^B
S_\gamma^C G_{\alpha\beta\gamma} ( \mathbf{r}_{\alpha,\beta} \times  \mathbf{r}_{\alpha,\gamma})
\label{eq:app_force}
\end{equation}
where $ \mathbf{r}_{\alpha,\beta}=\mathbf{r}_{\alpha}-\mathbf{r}_{\beta}$ and
\begin{equation}
G_{\alpha\beta\gamma}=G\left(| \mathbf{r}_{\alpha,\beta}|,| \mathbf{r}_{\alpha,\gamma}|\right)
\end{equation}
is an isotropic three-body interaction kernel that depends on the distances between the particles. Note we take the sum in Eq.~\ref{eq:app_force} to exclude the same particles. 

The presence of the Levi--Civita tensor in species space enforces cyclic coupling among the three species and guarantees that the interaction vanishes whenever two species coincide. Unlike forces derived from a pair potential, Eq.~\ref{eq:app_force} is generally nonreciprocal and nonconservative. It therefore provides a microscopic source of nonequilibrium parity breaking while preserving spatial isotropy.

\subsection{Microscopic Density Fields}
For each species we define the microscopic density
\begin{equation}
\hat\rho^A(\mathbf r,t)= \sum_\alpha S_\alpha^A \delta  [\mathbf r-\mathbf r_\alpha(t)].
\label{eq:app_density}
\end{equation}
whose equation of motion can be deduced by
applying Ito's lemma. This yields
\begin{equation}
d\hat\rho^A = \sum_\alpha S_\alpha^A \left ( \frac{\partial\hat\rho^A} {\partial r_{\alpha,i}} dr_{\alpha,i} + \frac12  dr_{\alpha,i} dr_{\alpha,j} \frac{\partial^2\hat\rho^A} {\partial r_{\alpha,i}\partial r_{\alpha,j}} \right )
\end{equation}
Using 
\begin{equation}
dr_{\alpha,i} = \mu F_{\alpha,i}dt + \sqrt{2D}dW_{\alpha,i},
\end{equation}
where $dW_{\alpha,i}$ is the stochastic increment associated with the noise $\eta$, together with
\begin{equation}
dW_{\alpha,i}dW_{\beta,j} = \delta_{\alpha\beta} \delta_{ij} dt,
\end{equation}
yields
\begin{equation}
dr_{\alpha,i}dr_{\alpha,j}=2D\delta_{ij}dt + \mathcal O(dt^2).
\end{equation}
Substituting these expressions and using
\begin{equation}
\partial_{r_{\alpha,i}} \delta(\mathbf r-\mathbf r_\alpha)= -\partial_i \delta(\mathbf r-\mathbf r_\alpha),
\end{equation}
one obtains the exact Dean equation
\begin{equation}
\partial_t\hat\rho^A = -\partial_i \hat J_i^A,
\label{eq:app_dean}
\end{equation}
with microscopic current
\begin{equation}
\hat J_i^A= -D\partial_i\hat\rho^A + \hat J_{i,\mathrm{int}}^A + \hat J_{i,\mathrm{noise}}^A 
\label{eq:app_current}
\end{equation}
which includes a standard diffusive term, $-D\partial_i\hat\rho^A $, a term from the interactions $\hat J_{i,\mathrm{int}}^A $
\begin{equation}
\hat J_{i,\mathrm{int}}^A= \mu \sum_\alpha S_\alpha^A F_{\alpha,i} \delta(\mathbf r-\mathbf r_\alpha)
\label{eq:app_interaction_current}
\end{equation}
 and a stochastic term that arises from the noise $\hat J_{i,\mathrm{noise}}^A$, and upon averaging will drop out. 

\subsection{Interaction Current in Density Form}

We now rewrite the microscopic interaction current entirely in terms of microscopic density fields. Substituting the force law Eq.~\ref{eq:app_force} into Eq.~\ref{eq:app_interaction_current} yields
\begin{align}
\hat J_{i,\mathrm{int}}^A(\mathbf r)= &\mu \lambda \varepsilon_{ijk} \varepsilon^{BCD} \sum_{\alpha,\beta,\gamma} S_\alpha^A S_\alpha^B S_\beta^C S_\gamma^D
\\
& G_{\alpha\beta\gamma} (r_{\alpha,j}-r_{\beta,j}) (r_{\alpha,k}-r_{\gamma,k}) \delta(\mathbf r-\mathbf r_\alpha),\nonumber
\end{align}
where sum over $\alpha$, $\beta$, and $\gamma$ is restricted to distinct particles. Using the projector identity the species indices simplify to
\begin{align}
\hat J_{i,\mathrm{int}}^A(\mathbf r) =& \mu \lambda \varepsilon_{ijk} \varepsilon^{ABC} \sum_{\alpha,\beta,\gamma} S_\alpha^A S_\beta^B S_\gamma^C\\
& G_{\alpha\beta\gamma} (r_{\alpha,j}-r_{\beta,j}) (r_{\alpha,k}-r_{\gamma,k}) \delta(\mathbf r-\mathbf r_\alpha) \nonumber
\end{align}
To express this quantity in continuum form we introduce the microscopic densities $\hat\rho^A(\mathbf r)$. The particle sums may then be rewritten as integrals over microscopic density fields. Introducing relative coordinates
and exploiting translational invariance of the kernel gives
\begin{align}
\label{eq:app_J3}
\hat J_{i,\mathrm{int}}^A(\mathbf r)=& \mu \lambda \varepsilon_{ijk} \varepsilon^{ABC} \int d\mathbf{r}' \int d \mathbf{r}''  G(\mathbf r',\mathbf r'')
\nonumber\\
& r_j' r''_k  \hat\rho^A(\mathbf r) \hat\rho^B(\mathbf r+\mathbf r') \hat\rho^C(\mathbf r+\mathbf r''),
\end{align}
where $G(\mathbf r',\mathbf r'')$ is the isotropic three-body kernel expressed in relative coordinates.

Equation~\ref{eq:app_J3} is the exact microscopic current associated with the three-body interaction. Several important features are immediately apparent.  First, the interaction couples three distinct species through the antisymmetric tensor $\varepsilon^{ABC}$. Consequently, the current vanishes identically whenever fewer than three species are present. Second, parity breaking enters exclusively through the spatial Levi--Civita tensor $\varepsilon_{ijk}$. The interaction remains isotropic because the kernel depends only on rotational invariants. Finally, Eq.~\ref{eq:app_J3} already exhibits the tensor structure responsible for odd transport. The current is constructed from an antisymmetric contraction of two displacement vectors, suggesting that the coarse-grained theory will involve an antisymmetric combination of density gradients.

The remaining task is therefore to determine the long-wavelength limit of Eq.~\ref{eq:app_J3}. To do so we first average over microscopic fluctuations and then perform a systematic gradient expansion.

\subsection{Ensemble Averaging and Closure}

To obtain a hydrodynamic description we average the microscopic current Eq.~\ref{eq:app_J3} over realizations of the stochastic dynamics. Defining
\begin{equation}
J_i^A(\mathbf r)= \left< \hat J_i^A(\mathbf r) \right>,
\end{equation}
we find
\begin{align}
J_{i,\mathrm{int}}^A(\mathbf r)= \mu \lambda \varepsilon_{ijk} \varepsilon^{ABC} \int d \mathbf{r}\int d \mathbf{r}'' G(\mathbf r',\mathbf r'')
\nonumber\\
 r_j' r''_k \rho^{ABC}( \mathbf r, \mathbf r+\mathbf r', \mathbf r+\mathbf r''),
\label{eq:app_avg_current}
\end{align}
where
\begin{align}
\rho^{ABC}( \mathbf r, \mathbf r+\mathbf r', \mathbf r+\mathbf r'')= \Big< \hat\rho^A(\mathbf r) \hat\rho^B(\mathbf r+\mathbf r') \hat\rho^C(\mathbf r+\mathbf r'')\Big>
\label{eq:threebodycorr}
\end{align}
is the equal-time three-body correlation function. Equation~\ref{eq:app_avg_current} is exact. However, it is not closed because the current depends on a three-body correlation function whose evolution involves higher-order correlations. As in conventional kinetic theories, the hydrodynamic description therefore requires a closure approximation. The constitutive law derived in the main text describes transport on length scales much larger than the range of the interaction kernel. Let $\ell$ denote the characteristic interaction range of $G(\mathbf r,\mathbf r')$ and let $L$ denote the characteristic length scale over which the densities vary. We assume $\ell \ll L$.

Under this condition the density fields vary only weakly across the support of the kernel and a local gradient expansion becomes possible. To leading order, correlations generated by the interaction remain short ranged. We therefore approximate the three-body density by
\begin{equation}
\rho^{ABC} ( \mathbf r, \mathbf r+\mathbf r', \mathbf r+\mathbf r'') \simeq \rho^A(\mathbf r) \rho^B(\mathbf r+\mathbf r') \rho^C(\mathbf r+\mathbf r''),
\label{eq:mfclosure}
\end{equation}
where $\rho^A(\mathbf r)= \left< \hat\rho^A(\mathbf r) \right>$ denotes the averaged density field. Equation~\ref{eq:mfclosure} corresponds to a mean-field factorization of the three-body correlation function. Corrections to this approximation generate higher-order gradient terms and renormalizations of the effective transport coefficient but do not alter the symmetry structure of the resulting constitutive law.
Substituting Eq.~\ref{eq:mfclosure} into Eq.~\ref{eq:app_avg_current} yields
\begin{align}
J_{i,\mathrm{int}}^A(\mathbf r)= \mu \lambda \varepsilon_{ijk} \varepsilon^{ABC} \rho^A(\mathbf r) \int d \mathbf{r}' \int d\mathbf{r}' G(\mathbf r',\mathbf r'')
\nonumber\\
 r_j' r''_k \rho^B(\mathbf r+\mathbf r') \rho^C(\mathbf r+\mathbf r'').
\label{eq:app_mf_current}
\end{align}
Equation~\ref{eq:app_mf_current} provides a closed nonlocal constitutive relation for the interaction current. The remaining task is to determine its long-wavelength form. This is accomplished by systematically expanding the densities about the point $\mathbf r$.

\subsection{Gradient Expansion and Emergence of Odd Diffusion}
We now evaluate the long-wavelength limit of Eq.~\ref{eq:app_mf_current}. Substituting the gradient expansions
\begin{align}
\rho^B(\mathbf r+\mathbf r') &= \rho^B+r_a'\partial_a\rho^B + \frac12 r_a' r_b' \partial_a\partial_b\rho^B+\cdots,
\\
\rho^C(\mathbf r+\mathbf r') &=\rho^C + r'_c\partial_c\rho^C+\frac12r'_c r'_d \partial_c\partial_d\rho^C +\cdots ,
\end{align}
into Eq.~\ref{eq:app_mf_current} generates a hierarchy of terms containing increasing numbers of gradients. To determine the leading contribution, we examine these terms order by order by writing
\begin{equation}
J_{i,\mathrm{int}}^A(\mathbf r) = J_{i,\mathrm{int}}^{A,(0)} + J_{i,\mathrm{int}}^{A,(1)} +J_{i,\mathrm{int}}^{A,(2)} +\dots
\end{equation}
where the super script denotes the order of the term.

The zeroth-order contribution is
\begin{align}
J_{i,\mathrm{int}}^{A,(0)}= \mu \lambda \varepsilon_{ijk} \varepsilon^{ABC} \rho^A\rho^B\rho^C  \int d \mathbf{r}',d \mathbf{r}'' G(\mathbf r',\mathbf r'') r'_j r''_k .
\end{align}
Rotational invariance implies
\begin{equation}
\int d \mathbf{r}',d \mathbf{r}'' G(\mathbf r',\mathbf r'') r'_j r''_k = M_2 \delta_{jk},
\end{equation}
for some scalar coefficient $M_2$. Contracting with the Levi--Civita tensor yields
\begin{equation}
\varepsilon_{ijk}\delta_{jk}=0,
\end{equation}
and therefore
\begin{equation}
J_{i,\mathrm{int}}^{A,(0)}=0
\end{equation}
or that the zeroth-order contribution vanishes identically.

Terms linear in gradients take the form
\begin{align}
J_{i,\mathrm{int}}^{A,(1)}= \mu \lambda \varepsilon_{ijk} \varepsilon^{ABC} \rho^A  \Big[ (\partial_a\rho^B)\rho^C  I_{jka}+ \rho^B(\partial_a\rho^C)  I'_{jka} \Big],
\end{align}
where
\begin{align}
I_{jka} &= \int d \mathbf{r}'\,d \mathbf{r}'' G(\mathbf r',\mathbf r'') r'_j r''_k r'_a , \\
I'_{jka} &= \int d \mathbf{r}'\,d \mathbf{r}'' G(\mathbf r',\mathbf r'') r'_j r''_k r''_a .
\end{align}
These integrals contain an odd number of displacement vectors. Since the kernel is isotropic and inversion symmetric,
\begin{equation}
G(\mathbf r,\mathbf r')=G(-\mathbf r,-\mathbf r'),
\end{equation}
both tensors vanish,
\begin{equation}
I_{jka}=I'_{jka}=0.
\end{equation}
and consequently,
\begin{equation}
J_{i,\mathrm{int}}^{A,(1)}=0.
\end{equation}
the leading contribution therefore appears only at second order in gradients.

The first nonvanishing term arises from the product of the two first-order gradient corrections,
\begin{align}
J_{i,\mathrm{int}}^{A,(2)}=\mu \lambda\varepsilon_{ijk}\varepsilon^{ABC}\rho^A(\partial_a\rho^B)(\partial_b\rho^C) M_{jkab},
\label{eq:secondordercurrent}
\end{align}
with
\begin{equation}
M_{jkab}= \int d \mathbf{r}' d \mathbf{r}'' G(\mathbf r',\mathbf r'') r'_j r''_k r'_a r''_b .
\label{eq:Mtensor}
\end{equation}
where the tensor $M_{jkab}$ is a fourth-rank isotropic tensor. Rotational invariance therefore implies the decomposition
\begin{align}
M_{jkab}=& A \delta_{ja}\delta_{kb}+B\delta_{jb}\delta_{ka}+ C\delta_{jk}\delta_{ab},
\label{eq:isotropicdecomp}
\end{align}
where $A$, $B$, and $C$ are scalar moments of the interaction kernel. Substituting Eq.~\ref{eq:isotropicdecomp} into Eq.~\ref{eq:secondordercurrent} gives
\begin{align}
J_{i,\mathrm{int}}^{A,(2)}= \mu \lambda \varepsilon^{ABC} \rho^A \Big[ A\varepsilon_{ijk}\delta_{ja}\delta_{kb}
\nonumber\\
+B\varepsilon_{ijk} \delta_{jb}\delta_{ka}+ C\varepsilon_{ijk} \delta_{jk}\delta_{ab} \Big] (\partial_a\rho^B)(\partial_b\rho^C).
\end{align}
where the final term vanishes immediately because
\begin{equation}
\varepsilon_{ijk}\delta_{jk}=0.
\end{equation}
The remaining two terms combine to give
\begin{align}
J_{i,\mathrm{int}}^{A,(2)}=
\mu \lambda (A-B) \varepsilon_{iab}\varepsilon^{ABC} \rho^A  (\partial_a\rho^B) (\partial_b\rho^C).
\label{eq:preodd}
\end{align}
Defining the odd parameter as $\omega=\mu \lambda (A-B)\rho^A$ , we can rewrite the current in a more compact form as 
\begin{align}
J_{i,\mathrm{int}}^{A,(2)}=
\omega \varepsilon_{iab}\varepsilon^{ABC}  (\partial_a\rho^B) (\partial_b\rho^C),
\label{eq:preodd}
\end{align}
which is the leading parity-odd contribution generated by the microscopic interaction and already of the form of the law proposed in the main text.

\subsection{Evaluation of the Transport Coefficient}
The derivation above establishes that the leading parity-odd contribution to the current is controlled by the difference between the isotropic tensor coefficients A and B appearing in Eq.~\ref{eq:isotropicdecomp}. The coefficients A, B, and C are determined by contraction with appropriate combinations of Kronecker deltas. Defining
\begin{align}
T_1 &= M_{jkab}\delta_{ja}\delta_{kb},\\
T_2 &= M_{jkab}\delta_{jb}\delta_{ka},\\
T_3 &= M_{jkab}\delta_{jk}\delta_{ab},
\end{align}
one obtains the linear system
\begin{align}
T_1 &= 9A+3B+3C,\\
T_2 &= 3A+9B+3C, \\
T_3 &= 3A+3B+9C.
\end{align}
which solving gives
\begin{align}
A-B=\frac{T_1-T_2}{6}.
\label{eq:AminusB}
\end{align}
Substituting the definitions of $T_1$ and $T_2$ yields
\begin{align}
A-B= \frac16 \int d\mathbf{r} d\mathbf{r}' G(\mathbf r,\mathbf r') \Big[ |\mathbf r|^2|\mathbf r'|^2- (\mathbf r\cdot\mathbf r')^2 \Big].
\label{eq:AminusBmoment}
\end{align}
The quantity in brackets is nonnegative and may be written as $|\mathbf r|^2|\mathbf r'|^2- (\mathbf r\cdot\mathbf r')^2 =|\mathbf r \times\mathbf r'|^2 $. Consequently,
the odd transport coefficient therefore becomes
\begin{equation}
\omega= -\frac{\mu \lambda}{6} \rho^A(\mathbf{r}) \int d \mathbf{r} d\mathbf{r}' G(\mathbf r,\mathbf r') |\mathbf r\times\mathbf r'|^2 .
\label{eq:omegafinal}
\end{equation}
Equation~\ref{eq:omegafinal} provides a direct microscopic interpretation of the hydrodynamic transport coefficient. The magnitude of the odd response is controlled by the average oriented area spanned by pairs of interaction vectors within a three-body cluster. Collinear particle configurations do not contribute to $\omega$, whereas noncoplanar arrangements maximize the odd transport. Combining Eq.~\ref{eq:omegafinal} with the gradient expansion derived above finally yields the nonlinear constitutive law employed throughout the main text. The derivation therefore establishes a complete microscopic route from isotropic nonreciprocal three-body interactions to the nonlinear odd transport theory developed in this work.

Several immediate consequences follow. First, the coefficient vanishes if the interaction kernel factorizes in a manner that eliminates angular correlations between $\mathbf r$ and $\mathbf r'$. Odd transport therefore requires genuinely three-body geometric information. Second, $\omega$ is proportional to the interaction strength $\lambda$ and mobility $\mu$, as expected from dimensional analysis. Third, the sign of $\omega$ is determined by the sign convention chosen in the microscopic force law. Reversing the cyclic ordering of the species labels changes
\begin{equation}
\varepsilon^{ABC} = -\varepsilon^{CBA},
\end{equation}
and therefore reverses the handedness of the resulting macroscopic currents.

\bibliography{references2B}

\begin{thebibliography}{37}%
\makeatletter
\providecommand \@ifxundefined [1]{%
 \@ifx{#1\undefined}
}%
\providecommand \@ifnum [1]{%
 \ifnum #1\expandafter \@firstoftwo
 \else \expandafter \@secondoftwo
 \fi
}%
\providecommand \@ifx [1]{%
 \ifx #1\expandafter \@firstoftwo
 \else \expandafter \@secondoftwo
 \fi
}%
\providecommand \natexlab [1]{#1}%
\providecommand \enquote  [1]{``#1''}%
\providecommand \bibnamefont  [1]{#1}%
\providecommand \bibfnamefont [1]{#1}%
\providecommand \citenamefont [1]{#1}%
\providecommand \href@noop [0]{\@secondoftwo}%
\providecommand \href [0]{\begingroup \@sanitize@url \@href}%
\providecommand \@href[1]{\@@startlink{#1}\@@href}%
\providecommand \@@href[1]{\endgroup#1\@@endlink}%
\providecommand \@sanitize@url [0]{\catcode `\\12\catcode `\$12\catcode
  `\&12\catcode `\#12\catcode `\^12\catcode `\_12\catcode `\%12\relax}%
\providecommand \@@startlink[1]{}%
\providecommand \@@endlink[0]{}%
\providecommand \url  [0]{\begingroup\@sanitize@url \@url }%
\providecommand \@url [1]{\endgroup\@href {#1}{\urlprefix }}%
\providecommand \urlprefix  [0]{URL }%
\providecommand \Eprint [0]{\href }%
\providecommand \doibase [0]{https://doi.org/}%
\providecommand \selectlanguage [0]{\@gobble}%
\providecommand \bibinfo  [0]{\@secondoftwo}%
\providecommand \bibfield  [0]{\@secondoftwo}%
\providecommand \translation [1]{[#1]}%
\providecommand \BibitemOpen [0]{}%
\providecommand \bibitemStop [0]{}%
\providecommand \bibitemNoStop [0]{.\EOS\space}%
\providecommand \EOS [0]{\spacefactor3000\relax}%
\providecommand \BibitemShut  [1]{\csname bibitem#1\endcsname}%
\let\auto@bib@innerbib\@empty
\bibitem [{\citenamefont {Scheibner}\ \emph {et~al.}(2021)\citenamefont
  {Scheibner}, \citenamefont {Irvine},\ and\ \citenamefont
  {Vitelli}}]{scheibner2021odd}%
  \BibitemOpen
  \bibfield  {author} {\bibinfo {author} {\bibfnamefont {C.}~\bibnamefont
  {Scheibner}}, \bibinfo {author} {\bibfnamefont {W.~T.~M.}\ \bibnamefont
  {Irvine}},\ and\ \bibinfo {author} {\bibfnamefont {V.}~\bibnamefont
  {Vitelli}},\ }\bibfield  {title} {\bibinfo {title} {Odd viscosity and odd
  elasticity},\ }\href {https://doi.org/10.1103/PhysRevLett.127.178001}
  {\bibfield  {journal} {\bibinfo  {journal} {Physical Review Letters}\
  }\textbf {\bibinfo {volume} {127}},\ \bibinfo {pages} {178001} (\bibinfo
  {year} {2021})}\BibitemShut {NoStop}%
\bibitem [{\citenamefont {Hargus}\ \emph {et~al.}(2021)\citenamefont {Hargus},
  \citenamefont {Epstein},\ and\ \citenamefont {Mandadapu}}]{hargus2021odd}%
  \BibitemOpen
  \bibfield  {author} {\bibinfo {author} {\bibfnamefont {C.}~\bibnamefont
  {Hargus}}, \bibinfo {author} {\bibfnamefont {J.~M.}\ \bibnamefont
  {Epstein}},\ and\ \bibinfo {author} {\bibfnamefont {K.~K.}\ \bibnamefont
  {Mandadapu}},\ }\bibfield  {title} {\bibinfo {title} {Odd diffusivity of
  chiral random motion},\ }\href
  {https://doi.org/10.1103/PhysRevLett.127.178001} {\bibfield  {journal}
  {\bibinfo  {journal} {Physical Review Letters}\ }\textbf {\bibinfo {volume}
  {127}},\ \bibinfo {pages} {178001} (\bibinfo {year} {2021})}\BibitemShut
  {NoStop}%
\bibitem [{\citenamefont {Poggioli}\ and\ \citenamefont
  {Limmer}(2023{\natexlab{a}})}]{poggioli2023odd}%
  \BibitemOpen
  \bibfield  {author} {\bibinfo {author} {\bibfnamefont {A.~R.}\ \bibnamefont
  {Poggioli}}\ and\ \bibinfo {author} {\bibfnamefont {D.~T.}\ \bibnamefont
  {Limmer}},\ }\bibfield  {title} {\bibinfo {title} {Odd mobility of a passive
  tracer in a chiral active fluid},\ }\href@noop {} {\bibfield  {journal}
  {\bibinfo  {journal} {Physical Review Letters}\ }\textbf {\bibinfo {volume}
  {130}},\ \bibinfo {pages} {158201} (\bibinfo {year}
  {2023}{\natexlab{a}})}\BibitemShut {NoStop}%
\bibitem [{\citenamefont {Banerjee}\ \emph {et~al.}(2017)\citenamefont
  {Banerjee}, \citenamefont {Souslov}, \citenamefont {Abanov},\ and\
  \citenamefont {Vitelli}}]{banerjee2017odd}%
  \BibitemOpen
  \bibfield  {author} {\bibinfo {author} {\bibfnamefont {D.}~\bibnamefont
  {Banerjee}}, \bibinfo {author} {\bibfnamefont {A.}~\bibnamefont {Souslov}},
  \bibinfo {author} {\bibfnamefont {A.~G.}\ \bibnamefont {Abanov}},\ and\
  \bibinfo {author} {\bibfnamefont {V.}~\bibnamefont {Vitelli}},\ }\bibfield
  {title} {\bibinfo {title} {Odd viscosity in chiral active fluids},\
  }\href@noop {} {\bibfield  {journal} {\bibinfo  {journal} {Nature
  communications}\ }\textbf {\bibinfo {volume} {8}},\ \bibinfo {pages} {1573}
  (\bibinfo {year} {2017})}\BibitemShut {NoStop}%
\bibitem [{\citenamefont {Markovich}\ and\ \citenamefont
  {Lubensky}(2024)}]{markovich2024nonreciprocity}%
  \BibitemOpen
  \bibfield  {author} {\bibinfo {author} {\bibfnamefont {T.}~\bibnamefont
  {Markovich}}\ and\ \bibinfo {author} {\bibfnamefont {T.~C.}\ \bibnamefont
  {Lubensky}},\ }\bibfield  {title} {\bibinfo {title} {Nonreciprocity and odd
  viscosity in chiral active fluids},\ }\href@noop {} {\bibfield  {journal}
  {\bibinfo  {journal} {Proceedings of the National Academy of Sciences}\
  }\textbf {\bibinfo {volume} {121}},\ \bibinfo {pages} {e2219385121} (\bibinfo
  {year} {2024})}\BibitemShut {NoStop}%
\bibitem [{\citenamefont {Poggioli}\ and\ \citenamefont
  {Limmer}(2023{\natexlab{b}})}]{poggioli2023emergent}%
  \BibitemOpen
  \bibfield  {author} {\bibinfo {author} {\bibfnamefont {A.~R.}\ \bibnamefont
  {Poggioli}}\ and\ \bibinfo {author} {\bibfnamefont {D.~T.}\ \bibnamefont
  {Limmer}},\ }\bibfield  {title} {\bibinfo {title} {Emergent kelvin waves in
  chiral active matter},\ }\href@noop {} {\bibfield  {journal} {\bibinfo
  {journal} {arXiv:2306.14984}\ } (\bibinfo {year}
  {2023}{\natexlab{b}})}\BibitemShut {NoStop}%
\bibitem [{\citenamefont {Alsallom}\ and\ \citenamefont
  {Limmer}(2026)}]{alsallom2026origin}%
  \BibitemOpen
  \bibfield  {author} {\bibinfo {author} {\bibfnamefont {F.}~\bibnamefont
  {Alsallom}}\ and\ \bibinfo {author} {\bibfnamefont {D.~T.}\ \bibnamefont
  {Limmer}},\ }\bibfield  {title} {\bibinfo {title} {Origin of edge currents in
  chiral active liquids},\ }\href@noop {} {\bibfield  {journal} {\bibinfo
  {journal} {arXiv:2603.18159}\ } (\bibinfo {year} {2026})}\BibitemShut
  {NoStop}%
\bibitem [{\citenamefont {Yang}\ \emph {et~al.}(2020)\citenamefont {Yang},
  \citenamefont {Ren}, \citenamefont {Cheng},\ and\ \citenamefont
  {Zhang}}]{yang2020robust}%
  \BibitemOpen
  \bibfield  {author} {\bibinfo {author} {\bibfnamefont {X.}~\bibnamefont
  {Yang}}, \bibinfo {author} {\bibfnamefont {C.}~\bibnamefont {Ren}}, \bibinfo
  {author} {\bibfnamefont {K.}~\bibnamefont {Cheng}},\ and\ \bibinfo {author}
  {\bibfnamefont {H.}~\bibnamefont {Zhang}},\ }\bibfield  {title} {\bibinfo
  {title} {Robust boundary flow in chiral active fluid},\ }\href@noop {}
  {\bibfield  {journal} {\bibinfo  {journal} {Physical Review E}\ }\textbf
  {\bibinfo {volume} {101}},\ \bibinfo {pages} {022603} (\bibinfo {year}
  {2020})}\BibitemShut {NoStop}%
\bibitem [{\citenamefont {Siebers}\ \emph {et~al.}(2024)\citenamefont
  {Siebers}, \citenamefont {Bebon}, \citenamefont {Jayaram},\ and\
  \citenamefont {Speck}}]{siebers2024collective}%
  \BibitemOpen
  \bibfield  {author} {\bibinfo {author} {\bibfnamefont {F.}~\bibnamefont
  {Siebers}}, \bibinfo {author} {\bibfnamefont {R.}~\bibnamefont {Bebon}},
  \bibinfo {author} {\bibfnamefont {A.}~\bibnamefont {Jayaram}},\ and\ \bibinfo
  {author} {\bibfnamefont {T.}~\bibnamefont {Speck}},\ }\bibfield  {title}
  {\bibinfo {title} {Collective hall current in chiral active fluids: Coupling
  of phase and mass transport through traveling bands},\ }\href@noop {}
  {\bibfield  {journal} {\bibinfo  {journal} {Proceedings of the National
  Academy of Sciences}\ }\textbf {\bibinfo {volume} {121}},\ \bibinfo {pages}
  {e2320256121} (\bibinfo {year} {2024})}\BibitemShut {NoStop}%
\bibitem [{\citenamefont {Soni}\ \emph {et~al.}(2019)\citenamefont {Soni},
  \citenamefont {Bililign}, \citenamefont {Magkiriadou}, \citenamefont
  {Sacanna}, \citenamefont {Bartolo}, \citenamefont {Shelley},\ and\
  \citenamefont {Irvine}}]{soni2019odd}%
  \BibitemOpen
  \bibfield  {author} {\bibinfo {author} {\bibfnamefont {V.}~\bibnamefont
  {Soni}}, \bibinfo {author} {\bibfnamefont {E.~S.}\ \bibnamefont {Bililign}},
  \bibinfo {author} {\bibfnamefont {S.}~\bibnamefont {Magkiriadou}}, \bibinfo
  {author} {\bibfnamefont {S.}~\bibnamefont {Sacanna}}, \bibinfo {author}
  {\bibfnamefont {D.}~\bibnamefont {Bartolo}}, \bibinfo {author} {\bibfnamefont
  {M.~J.}\ \bibnamefont {Shelley}},\ and\ \bibinfo {author} {\bibfnamefont
  {W.~T.}\ \bibnamefont {Irvine}},\ }\bibfield  {title} {\bibinfo {title} {The
  odd free surface flows of a colloidal chiral fluid},\ }\href
  {https://doi.org/10.1038/s41567-019-0603-8} {\bibfield  {journal} {\bibinfo
  {journal} {Nature Physics}\ }\textbf {\bibinfo {volume} {15}},\ \bibinfo
  {pages} {1188} (\bibinfo {year} {2019})}\BibitemShut {NoStop}%
\bibitem [{\citenamefont {Fruchart}\ \emph {et~al.}(2023)\citenamefont
  {Fruchart}, \citenamefont {Scheibner},\ and\ \citenamefont
  {Vitelli}}]{fruchart2023odd}%
  \BibitemOpen
  \bibfield  {author} {\bibinfo {author} {\bibfnamefont {M.}~\bibnamefont
  {Fruchart}}, \bibinfo {author} {\bibfnamefont {C.}~\bibnamefont
  {Scheibner}},\ and\ \bibinfo {author} {\bibfnamefont {V.}~\bibnamefont
  {Vitelli}},\ }\bibfield  {title} {\bibinfo {title} {Odd viscosity and odd
  elasticity},\ }\href@noop {} {\bibfield  {journal} {\bibinfo  {journal}
  {Annual Review of Condensed Matter Physics}\ }\textbf {\bibinfo {volume}
  {14}},\ \bibinfo {pages} {471} (\bibinfo {year} {2023})}\BibitemShut
  {NoStop}%
\bibitem [{\citenamefont {Metzger}\ \emph {et~al.}(2026)\citenamefont
  {Metzger}, \citenamefont {Hargus}, \citenamefont {Tailleur},\ and\
  \citenamefont {van Wijland}}]{metzger2026equation}%
  \BibitemOpen
  \bibfield  {author} {\bibinfo {author} {\bibfnamefont {J.}~\bibnamefont
  {Metzger}}, \bibinfo {author} {\bibfnamefont {C.}~\bibnamefont {Hargus}},
  \bibinfo {author} {\bibfnamefont {J.}~\bibnamefont {Tailleur}},\ and\
  \bibinfo {author} {\bibfnamefont {F.}~\bibnamefont {van Wijland}},\
  }\bibfield  {title} {\bibinfo {title} {Equation of state for the edge flow of
  chiral colloidal fluids},\ }\href@noop {} {\bibfield  {journal} {\bibinfo
  {journal} {arXiv:2604.18708}\ } (\bibinfo {year} {2026})}\BibitemShut
  {NoStop}%
\bibitem [{\citenamefont {Kalz}\ \emph {et~al.}(2022)\citenamefont {Kalz},
  \citenamefont {Vuijk}, \citenamefont {Abdoli}, \citenamefont {Sommer},
  \citenamefont {L{\"o}wen},\ and\ \citenamefont
  {Sharma}}]{kalz2022collisions}%
  \BibitemOpen
  \bibfield  {author} {\bibinfo {author} {\bibfnamefont {E.}~\bibnamefont
  {Kalz}}, \bibinfo {author} {\bibfnamefont {H.~D.}\ \bibnamefont {Vuijk}},
  \bibinfo {author} {\bibfnamefont {I.}~\bibnamefont {Abdoli}}, \bibinfo
  {author} {\bibfnamefont {J.-U.}\ \bibnamefont {Sommer}}, \bibinfo {author}
  {\bibfnamefont {H.}~\bibnamefont {L{\"o}wen}},\ and\ \bibinfo {author}
  {\bibfnamefont {A.}~\bibnamefont {Sharma}},\ }\bibfield  {title} {\bibinfo
  {title} {Collisions enhance self-diffusion in odd-diffusive systems},\
  }\href@noop {} {\bibfield  {journal} {\bibinfo  {journal} {Physical Review
  Letters}\ }\textbf {\bibinfo {volume} {129}},\ \bibinfo {pages} {090601}
  (\bibinfo {year} {2022})}\BibitemShut {NoStop}%
\bibitem [{\citenamefont {Hargus}\ \emph
  {et~al.}(2025{\natexlab{a}})\citenamefont {Hargus}, \citenamefont {Ghimenti},
  \citenamefont {Tailleur},\ and\ \citenamefont {van Wijland}}]{hargus2025odd}%
  \BibitemOpen
  \bibfield  {author} {\bibinfo {author} {\bibfnamefont {C.}~\bibnamefont
  {Hargus}}, \bibinfo {author} {\bibfnamefont {F.}~\bibnamefont {Ghimenti}},
  \bibinfo {author} {\bibfnamefont {J.}~\bibnamefont {Tailleur}},\ and\
  \bibinfo {author} {\bibfnamefont {F.}~\bibnamefont {van Wijland}},\
  }\bibfield  {title} {\bibinfo {title} {Odd dynamics of passive objects in a
  chiral active bath},\ }\href@noop {} {\bibfield  {journal} {\bibinfo
  {journal} {Physical Review Letters}\ }\textbf {\bibinfo {volume} {135}},\
  \bibinfo {pages} {167102} (\bibinfo {year} {2025}{\natexlab{a}})}\BibitemShut
  {NoStop}%
\bibitem [{\citenamefont {Langford}\ and\ \citenamefont
  {Omar}(2025)}]{langford2025phase}%
  \BibitemOpen
  \bibfield  {author} {\bibinfo {author} {\bibfnamefont {L.}~\bibnamefont
  {Langford}}\ and\ \bibinfo {author} {\bibfnamefont {A.~K.}\ \bibnamefont
  {Omar}},\ }\bibfield  {title} {\bibinfo {title} {Phase separation,
  capillarity, and odd-surface flows in chiral active matter},\ }\href@noop {}
  {\bibfield  {journal} {\bibinfo  {journal} {Physical Review Letters}\
  }\textbf {\bibinfo {volume} {134}},\ \bibinfo {pages} {068301} (\bibinfo
  {year} {2025})}\BibitemShut {NoStop}%
\bibitem [{\citenamefont {Hargus}\ \emph
  {et~al.}(2025{\natexlab{b}})\citenamefont {Hargus}, \citenamefont
  {Deshpande}, \citenamefont {Omar},\ and\ \citenamefont
  {Mandadapu}}]{hargus2025flux}%
  \BibitemOpen
  \bibfield  {author} {\bibinfo {author} {\bibfnamefont {C.}~\bibnamefont
  {Hargus}}, \bibinfo {author} {\bibfnamefont {A.}~\bibnamefont {Deshpande}},
  \bibinfo {author} {\bibfnamefont {A.~K.}\ \bibnamefont {Omar}},\ and\
  \bibinfo {author} {\bibfnamefont {K.~K.}\ \bibnamefont {Mandadapu}},\
  }\bibfield  {title} {\bibinfo {title} {Flux hypothesis for odd transport
  phenomena},\ }\href@noop {} {\bibfield  {journal} {\bibinfo  {journal}
  {Physical Review Letters}\ }\textbf {\bibinfo {volume} {134}},\ \bibinfo
  {pages} {097105} (\bibinfo {year} {2025}{\natexlab{b}})}\BibitemShut
  {NoStop}%
\bibitem [{\citenamefont {Vega~Reyes}\ \emph {et~al.}(2022)\citenamefont
  {Vega~Reyes}, \citenamefont {L{\'o}pez-Casta{\~n}o},\ and\ \citenamefont
  {Rodr{\'\i}guez-Rivas}}]{vega2022diffusive}%
  \BibitemOpen
  \bibfield  {author} {\bibinfo {author} {\bibfnamefont {F.}~\bibnamefont
  {Vega~Reyes}}, \bibinfo {author} {\bibfnamefont {M.~A.}\ \bibnamefont
  {L{\'o}pez-Casta{\~n}o}},\ and\ \bibinfo {author} {\bibfnamefont
  {{\'A}.}~\bibnamefont {Rodr{\'\i}guez-Rivas}},\ }\bibfield  {title} {\bibinfo
  {title} {Diffusive regimes in a two-dimensional chiral fluid},\ }\href@noop
  {} {\bibfield  {journal} {\bibinfo  {journal} {Communications Physics}\
  }\textbf {\bibinfo {volume} {5}},\ \bibinfo {pages} {256} (\bibinfo {year}
  {2022})}\BibitemShut {NoStop}%
\bibitem [{\citenamefont {von Klitzing}\ \emph {et~al.}(2020)\citenamefont {von
  Klitzing}, \citenamefont {Chakraborty}, \citenamefont {Kim}, \citenamefont
  {Madhavan}, \citenamefont {Dai}, \citenamefont {McIver}, \citenamefont
  {Tokura}, \citenamefont {Savary}, \citenamefont {Smirnova}, \citenamefont
  {Rey} \emph {et~al.}}]{von202040}%
  \BibitemOpen
  \bibfield  {author} {\bibinfo {author} {\bibfnamefont {K.}~\bibnamefont {von
  Klitzing}}, \bibinfo {author} {\bibfnamefont {T.}~\bibnamefont
  {Chakraborty}}, \bibinfo {author} {\bibfnamefont {P.}~\bibnamefont {Kim}},
  \bibinfo {author} {\bibfnamefont {V.}~\bibnamefont {Madhavan}}, \bibinfo
  {author} {\bibfnamefont {X.}~\bibnamefont {Dai}}, \bibinfo {author}
  {\bibfnamefont {J.}~\bibnamefont {McIver}}, \bibinfo {author} {\bibfnamefont
  {Y.}~\bibnamefont {Tokura}}, \bibinfo {author} {\bibfnamefont
  {L.}~\bibnamefont {Savary}}, \bibinfo {author} {\bibfnamefont
  {D.}~\bibnamefont {Smirnova}}, \bibinfo {author} {\bibfnamefont {A.~M.}\
  \bibnamefont {Rey}}, \emph {et~al.},\ }\bibfield  {title} {\bibinfo {title}
  {40 years of the quantum hall effect},\ }\href@noop {} {\bibfield  {journal}
  {\bibinfo  {journal} {Nature Reviews Physics}\ }\textbf {\bibinfo {volume}
  {2}},\ \bibinfo {pages} {397} (\bibinfo {year} {2020})}\BibitemShut {NoStop}%
\bibitem [{\citenamefont {Illien}(2025)}]{illien2025dean}%
  \BibitemOpen
  \bibfield  {author} {\bibinfo {author} {\bibfnamefont {P.}~\bibnamefont
  {Illien}},\ }\bibfield  {title} {\bibinfo {title} {The dean--kawasaki
  equation and stochastic density functional theory},\ }\href@noop {}
  {\bibfield  {journal} {\bibinfo  {journal} {Reports on Progress in Physics}\
  }\textbf {\bibinfo {volume} {88}},\ \bibinfo {pages} {086601} (\bibinfo
  {year} {2025})}\BibitemShut {NoStop}%
\bibitem [{\citenamefont {Irving}\ and\ \citenamefont
  {Kirkwood}(1950)}]{irving1950statistical}%
  \BibitemOpen
  \bibfield  {author} {\bibinfo {author} {\bibfnamefont {J.}~\bibnamefont
  {Irving}}\ and\ \bibinfo {author} {\bibfnamefont {J.~G.}\ \bibnamefont
  {Kirkwood}},\ }\bibfield  {title} {\bibinfo {title} {The statistical
  mechanical theory of transport processes. iv. the equations of
  hydrodynamics},\ }\href@noop {} {\bibfield  {journal} {\bibinfo  {journal}
  {The Journal of chemical physics}\ }\textbf {\bibinfo {volume} {18}},\
  \bibinfo {pages} {817} (\bibinfo {year} {1950})}\BibitemShut {NoStop}%
\bibitem [{\citenamefont {Kawasaki}(1994)}]{kawasaki1994stochastic}%
  \BibitemOpen
  \bibfield  {author} {\bibinfo {author} {\bibfnamefont {K.}~\bibnamefont
  {Kawasaki}},\ }\bibfield  {title} {\bibinfo {title} {Stochastic model of slow
  dynamics in supercooled liquids and dense colloidal suspensions},\
  }\href@noop {} {\bibfield  {journal} {\bibinfo  {journal} {Physica A:
  Statistical Mechanics and its Applications}\ }\textbf {\bibinfo {volume}
  {208}},\ \bibinfo {pages} {35} (\bibinfo {year} {1994})}\BibitemShut
  {NoStop}%
\bibitem [{\citenamefont {Dean}(1996)}]{dean1996langevin}%
  \BibitemOpen
  \bibfield  {author} {\bibinfo {author} {\bibfnamefont {D.~S.}\ \bibnamefont
  {Dean}},\ }\bibfield  {title} {\bibinfo {title} {Langevin equation for the
  density of a system of interacting langevin processes},\ }\href@noop {}
  {\bibfield  {journal} {\bibinfo  {journal} {Journal of Physics A:
  Mathematical and General}\ }\textbf {\bibinfo {volume} {29}},\ \bibinfo
  {pages} {L613} (\bibinfo {year} {1996})}\BibitemShut {NoStop}%
\bibitem [{\citenamefont {Jeffreys}(1973)}]{jeffreys1973isotropic}%
  \BibitemOpen
  \bibfield  {author} {\bibinfo {author} {\bibfnamefont {H.}~\bibnamefont
  {Jeffreys}},\ }\bibfield  {title} {\bibinfo {title} {On isotropic tensors},\
  }in\ \href@noop {} {\emph {\bibinfo {booktitle} {Mathematical Proceedings of
  the Cambridge philosophical society}}},\ Vol.~\bibinfo {volume} {73}\
  (\bibinfo {organization} {Cambridge University Press},\ \bibinfo {year}
  {1973})\ pp.\ \bibinfo {pages} {173--176}\BibitemShut {NoStop}%
\bibitem [{\citenamefont {Epstein}\ and\ \citenamefont
  {Mandadapu}(2020)}]{epstein2020time}%
  \BibitemOpen
  \bibfield  {author} {\bibinfo {author} {\bibfnamefont {J.~M.}\ \bibnamefont
  {Epstein}}\ and\ \bibinfo {author} {\bibfnamefont {K.~K.}\ \bibnamefont
  {Mandadapu}},\ }\bibfield  {title} {\bibinfo {title} {Time-reversal symmetry
  breaking in two-dimensional nonequilibrium viscous fluids},\ }\href@noop {}
  {\bibfield  {journal} {\bibinfo  {journal} {Physical review E}\ }\textbf
  {\bibinfo {volume} {101}},\ \bibinfo {pages} {052614} (\bibinfo {year}
  {2020})}\BibitemShut {NoStop}%
\bibitem [{\citenamefont {Tong}(2025)}]{tong2025electromagnetism}%
  \BibitemOpen
  \bibfield  {author} {\bibinfo {author} {\bibfnamefont {D.}~\bibnamefont
  {Tong}},\ }\href@noop {} {\emph {\bibinfo {title} {Electromagnetism: Lectures
  on Theoretical Physics}}}\ (\bibinfo  {publisher} {Cambridge University
  Press},\ \bibinfo {year} {2025})\BibitemShut {NoStop}%
\bibitem [{\citenamefont {Landau}\ and\ \citenamefont
  {Lifshitz}(1987)}]{landau1987fluid}%
  \BibitemOpen
  \bibfield  {author} {\bibinfo {author} {\bibfnamefont {L.~D.}\ \bibnamefont
  {Landau}}\ and\ \bibinfo {author} {\bibfnamefont {E.~M.}\ \bibnamefont
  {Lifshitz}},\ }\href@noop {} {\emph {\bibinfo {title} {Fluid Mechanics}}},\
  Vol.~\bibinfo {volume} {6}\ (\bibinfo  {publisher} {Elsevier},\ \bibinfo
  {year} {1987})\BibitemShut {NoStop}%
\bibitem [{\citenamefont {Doering}\ and\ \citenamefont
  {Gibbon}(1995)}]{doering1995applied}%
  \BibitemOpen
  \bibfield  {author} {\bibinfo {author} {\bibfnamefont {C.~R.}\ \bibnamefont
  {Doering}}\ and\ \bibinfo {author} {\bibfnamefont {J.~D.}\ \bibnamefont
  {Gibbon}},\ }\href@noop {} {\emph {\bibinfo {title} {Applied Analysis of the
  Navier-Stokes Equations}}},\ \bibinfo {series} {Cambridge Texts in Applied
  Mathematics}, Vol.~\bibinfo {volume} {12}\ (\bibinfo  {publisher} {Cambridge
  University Press},\ \bibinfo {address} {Cambridge},\ \bibinfo {year}
  {1995})\BibitemShut {NoStop}%
\bibitem [{\citenamefont {Jim{\'e}nez}(2018)}]{jimenez2018coherent}%
  \BibitemOpen
  \bibfield  {author} {\bibinfo {author} {\bibfnamefont {J.}~\bibnamefont
  {Jim{\'e}nez}},\ }\bibfield  {title} {\bibinfo {title} {Coherent structures
  in wall-bounded turbulence},\ }\href@noop {} {\bibfield  {journal} {\bibinfo
  {journal} {Journal of Fluid Mechanics}\ }\textbf {\bibinfo {volume} {842}},\
  \bibinfo {pages} {P1} (\bibinfo {year} {2018})}\BibitemShut {NoStop}%
\bibitem [{\citenamefont {Buaria}\ \emph {et~al.}(2020)\citenamefont {Buaria},
  \citenamefont {Bodenschatz},\ and\ \citenamefont {Pumir}}]{buaria2020vortex}%
  \BibitemOpen
  \bibfield  {author} {\bibinfo {author} {\bibfnamefont {D.}~\bibnamefont
  {Buaria}}, \bibinfo {author} {\bibfnamefont {E.}~\bibnamefont
  {Bodenschatz}},\ and\ \bibinfo {author} {\bibfnamefont {A.}~\bibnamefont
  {Pumir}},\ }\bibfield  {title} {\bibinfo {title} {Vortex stretching and
  enstrophy production in high reynolds number turbulence},\ }\href@noop {}
  {\bibfield  {journal} {\bibinfo  {journal} {Physical Review Fluids}\ }\textbf
  {\bibinfo {volume} {5}},\ \bibinfo {pages} {104602} (\bibinfo {year}
  {2020})}\BibitemShut {NoStop}%
\bibitem [{\citenamefont {Limmer}(2024)}]{limmer2024statistical}%
  \BibitemOpen
  \bibfield  {author} {\bibinfo {author} {\bibfnamefont {D.~T.}\ \bibnamefont
  {Limmer}},\ }\href {https://doi.org/10.1093/oso/9780198919858.001.0001}
  {\emph {\bibinfo {title} {Statistical Mechanics and Stochastic
  Thermodynamics}}},\ Oxford Graduate Texts\ (\bibinfo  {publisher} {OUP
  Oxford},\ \bibinfo {address} {Oxford},\ \bibinfo {year} {2024})\BibitemShut
  {NoStop}%
\bibitem [{\citenamefont {Gao}\ and\ \citenamefont
  {Limmer}(2019)}]{gao2019nonlinear}%
  \BibitemOpen
  \bibfield  {author} {\bibinfo {author} {\bibfnamefont {C.~Y.}\ \bibnamefont
  {Gao}}\ and\ \bibinfo {author} {\bibfnamefont {D.~T.}\ \bibnamefont
  {Limmer}},\ }\bibfield  {title} {\bibinfo {title} {Nonlinear transport
  coefficients from large deviation functions},\ }\href@noop {} {\bibfield
  {journal} {\bibinfo  {journal} {The Journal of chemical physics}\ }\textbf
  {\bibinfo {volume} {151}} (\bibinfo {year} {2019})}\BibitemShut {NoStop}%
\bibitem [{\citenamefont {Limmer}\ \emph {et~al.}(2021)\citenamefont {Limmer},
  \citenamefont {Gao},\ and\ \citenamefont {Poggioli}}]{limmer2021large}%
  \BibitemOpen
  \bibfield  {author} {\bibinfo {author} {\bibfnamefont {D.~T.}\ \bibnamefont
  {Limmer}}, \bibinfo {author} {\bibfnamefont {C.~Y.}\ \bibnamefont {Gao}},\
  and\ \bibinfo {author} {\bibfnamefont {A.~R.}\ \bibnamefont {Poggioli}},\
  }\bibfield  {title} {\bibinfo {title} {A large deviation theory perspective
  on nanoscale transport phenomena},\ }\href@noop {} {\bibfield  {journal}
  {\bibinfo  {journal} {The European Physical Journal B}\ }\textbf {\bibinfo
  {volume} {94}},\ \bibinfo {pages} {145} (\bibinfo {year} {2021})}\BibitemShut
  {NoStop}%
\bibitem [{\citenamefont {Deshpande}\ \emph {et~al.}(2024)\citenamefont
  {Deshpande}, \citenamefont {Hargus}, \citenamefont {Shekhar},\ and\
  \citenamefont {Mandadapu}}]{deshpande2024odd}%
  \BibitemOpen
  \bibfield  {author} {\bibinfo {author} {\bibfnamefont {A.}~\bibnamefont
  {Deshpande}}, \bibinfo {author} {\bibfnamefont {C.}~\bibnamefont {Hargus}},
  \bibinfo {author} {\bibfnamefont {K.}~\bibnamefont {Shekhar}},\ and\ \bibinfo
  {author} {\bibfnamefont {K.~K.}\ \bibnamefont {Mandadapu}},\ }\bibfield
  {title} {\bibinfo {title} {Odd viscodiffusive fluids},\ }\href@noop {}
  {\bibfield  {journal} {\bibinfo  {journal} {arXiv:2411.04309}\ } (\bibinfo
  {year} {2024})}\BibitemShut {NoStop}%
\bibitem [{\citenamefont {Neubauer}\ \emph {et~al.}(2009)\citenamefont
  {Neubauer}, \citenamefont {Pfleiderer}, \citenamefont {Binz}, \citenamefont
  {Rosch}, \citenamefont {Ritz}, \citenamefont {Niklowitz},\ and\ \citenamefont
  {B{\"o}ni}}]{neubauer2009topological}%
  \BibitemOpen
  \bibfield  {author} {\bibinfo {author} {\bibfnamefont {A.}~\bibnamefont
  {Neubauer}}, \bibinfo {author} {\bibfnamefont {C.}~\bibnamefont
  {Pfleiderer}}, \bibinfo {author} {\bibfnamefont {B.}~\bibnamefont {Binz}},
  \bibinfo {author} {\bibfnamefont {A.}~\bibnamefont {Rosch}}, \bibinfo
  {author} {\bibfnamefont {R.}~\bibnamefont {Ritz}}, \bibinfo {author}
  {\bibfnamefont {P.~G.}\ \bibnamefont {Niklowitz}},\ and\ \bibinfo {author}
  {\bibfnamefont {P.}~\bibnamefont {B{\"o}ni}},\ }\href
  {https://arxiv.org/abs/0902.1933} {\bibinfo {title} {Topological hall effect
  in the a-phase of mnsi}} (\bibinfo {year} {2009}),\ \Eprint
  {https://arxiv.org/abs/0902.1933} {arXiv:0902.1933} \BibitemShut {NoStop}%
\bibitem [{\citenamefont {Tucci}\ \emph {et~al.}(2024)\citenamefont {Tucci},
  \citenamefont {Golestanian},\ and\ \citenamefont
  {Saha}}]{tucci2024nonreciprocal}%
  \BibitemOpen
  \bibfield  {author} {\bibinfo {author} {\bibfnamefont {G.}~\bibnamefont
  {Tucci}}, \bibinfo {author} {\bibfnamefont {R.}~\bibnamefont {Golestanian}},\
  and\ \bibinfo {author} {\bibfnamefont {S.}~\bibnamefont {Saha}},\ }\bibfield
  {title} {\bibinfo {title} {Nonreciprocal collective dynamics in a mixture of
  phoretic janus colloids},\ }\href@noop {} {\bibfield  {journal} {\bibinfo
  {journal} {New Journal of Physics}\ }\textbf {\bibinfo {volume} {26}},\
  \bibinfo {pages} {073006} (\bibinfo {year} {2024})}\BibitemShut {NoStop}%
\bibitem [{\citenamefont {Bililign}\ \emph {et~al.}(2022)\citenamefont
  {Bililign}, \citenamefont {Balboa~Usabiaga}, \citenamefont {Ganan},
  \citenamefont {Poncet}, \citenamefont {Soni}, \citenamefont {Magkiriadou},
  \citenamefont {Shelley}, \citenamefont {Bartolo},\ and\ \citenamefont
  {Irvine}}]{bililign2022motile}%
  \BibitemOpen
  \bibfield  {author} {\bibinfo {author} {\bibfnamefont {E.~S.}\ \bibnamefont
  {Bililign}}, \bibinfo {author} {\bibfnamefont {F.}~\bibnamefont
  {Balboa~Usabiaga}}, \bibinfo {author} {\bibfnamefont {Y.~A.}\ \bibnamefont
  {Ganan}}, \bibinfo {author} {\bibfnamefont {A.}~\bibnamefont {Poncet}},
  \bibinfo {author} {\bibfnamefont {V.}~\bibnamefont {Soni}}, \bibinfo {author}
  {\bibfnamefont {S.}~\bibnamefont {Magkiriadou}}, \bibinfo {author}
  {\bibfnamefont {M.~J.}\ \bibnamefont {Shelley}}, \bibinfo {author}
  {\bibfnamefont {D.}~\bibnamefont {Bartolo}},\ and\ \bibinfo {author}
  {\bibfnamefont {W.~T.~M.}\ \bibnamefont {Irvine}},\ }\bibfield  {title}
  {\bibinfo {title} {Motile dislocations knead odd crystals into whorls},\
  }\href@noop {} {\bibfield  {journal} {\bibinfo  {journal} {Nature Physics}\
  }\textbf {\bibinfo {volume} {18}},\ \bibinfo {pages} {212} (\bibinfo {year}
  {2022})}\BibitemShut {NoStop}%
\bibitem [{\citenamefont {Mecke}\ \emph {et~al.}(2023)\citenamefont {Mecke},
  \citenamefont {Gao}, \citenamefont {Ram{\'i}rez~Medina}, \citenamefont
  {Aarts}, \citenamefont {Gompper},\ and\ \citenamefont
  {Ripoll}}]{mecke2023simultaneous}%
  \BibitemOpen
  \bibfield  {author} {\bibinfo {author} {\bibfnamefont {J.}~\bibnamefont
  {Mecke}}, \bibinfo {author} {\bibfnamefont {Y.}~\bibnamefont {Gao}}, \bibinfo
  {author} {\bibfnamefont {C.~A.}\ \bibnamefont {Ram{\'i}rez~Medina}}, \bibinfo
  {author} {\bibfnamefont {D.~G. A.~L.}\ \bibnamefont {Aarts}}, \bibinfo
  {author} {\bibfnamefont {G.}~\bibnamefont {Gompper}},\ and\ \bibinfo {author}
  {\bibfnamefont {M.}~\bibnamefont {Ripoll}},\ }\bibfield  {title} {\bibinfo
  {title} {Simultaneous emergence of active turbulence and odd viscosity in a
  colloidal chiral active system},\ }\href@noop {} {\bibfield  {journal}
  {\bibinfo  {journal} {Communications Physics}\ }\textbf {\bibinfo {volume}
  {6}},\ \bibinfo {pages} {324} (\bibinfo {year} {2023})}\BibitemShut {NoStop}%
\end{thebibliography}%

\end{document}